# Cosmology Population Monte Carlo

CosmoPMC v1.2

User's manual

**Martin Kilbinger**
**Karim Benabed**
**Olivier Cappé**
**Jean Coupon**
**Jean-François Cardoso**
**Gersende Fort**
**Henry J. McCracken**
**Simon Prunet**
**Christian P. Robert**
**Darren Wraith**

# Contents

# 1. What is CosmoPMC?

name

CosmoPMC (Cosmology Population Monte Carlo) is a Bayesian sampling method to explore the likelihood of various cosmological probes. The sampling engine is implemented with the package pmclib. It is called Population Monte Carlo (PMC), which is a novel technique to sample from the posterior (Cappé et al. 2008). PMC is an adaptive importance sampling method which iteratively improves the proposal to approximate the posterior. This code has been introduced, tested and applied to various cosmology data sets in Wraith et al. (2009). Results on the Bayesian evidence using PMC are discussed in Kilbinger et al. (2010).

## 1.1. Importance sampling

One of the main goals in Bayesian inference is to obtain integrals of the form

$$\pi(f) = \int f(x)\pi(x)\mathrm{d}x \tag{1}$$

over the posterior distribution $\pi$ which depends on the $p$-dimensional parameter $x$, where $f$ is an arbitrary function with finite expectation under $\pi$. Of interest are for example the parameter mean ($f = \mathrm{id}$) or confidence regions $S$ with $f = \mathbf{1}_S$ being the indicator function of $S$. The Bayesian evidence $E$, used in model comparison techniques, is obtained by setting $f = 1$, but instead of $\pi$ using the unnormalised posterior $\pi' = L \cdot P$ in (1), with $L$ being the likelihood and $P$ the prior.

The evaluation of (1) is challenging because the posterior is in general not available analytically, and the parameter space can be high-dimensional. Monte-Carlo methods to approximate the above integrals consist in providing a sample $\{x_n\}_{n=1\ldots N}$ under $\pi$, and approximating (1) by the estimator

$$\hat{\pi}(f) = \frac{1}{N}\sum_{n=1}^{N} f(x_n). \tag{2}$$

Markov Chain Monte Carlo (MCMC) produces a Markov chain of points for which $\pi$ is the limiting distribution. The popular and widely-used package cosmomc (`http://cosmologist.info/cosmomc`; Lewis & Bridle 2002) implements MCMC exploration of the cosmological parameter space.

Importance sampling on the other hand uses the identity

$$\pi(f) = \int f(x)\pi(x)\mathrm{d}x = \int f(x)\frac{\pi(x)}{q(x)}q(x)\,\mathrm{d}x \tag{3}$$

where $q$ is any probability density function with support including the support of $\pi$. A sample $\{x_n\}$ under $q$ is then used to obtain the estimator

$$\hat{\pi}(f) = \frac{1}{N} \sum_{n=1}^{N} f(x_n)\, w_n; \qquad w_n = \frac{\pi(x_n)}{q(x_n)}. \tag{4}$$

The function $q$ is called the *proposal* or *importance function*, the quantities $w_n$ are the *importance weights*. Population Monte Carlo (PMC) produces a sequence $q^t$ of importance functions ($t = 1 \ldots T$) to approximate the posterior $\pi$. Details of this algorithm are discussed in Wraith et al. (2009).

The package CosmoPMC provides a C-code for sampling and exploring the cosmological parameter space using Population Monte Carlo. The code uses MPI to parallelize the calculation of the likelihood function. There is very little overhead and on a massive cluster the reduction in wall-clock time can be enormous. Included in the package are post-processing, plotting and various other analysis scripts and programs. It also provides a Markov Chain Monte-Carlo sampler.

### 1.2. This manual

This manual describes the code CosmoPMC, and can be obtained from `www.cosmopmc.info`. CosmoPMC is the cosmology interface to the Population Monte Carlo (PMC) engine pmclib. Documentation on the PMC library can be found at the same url. The cosmology module of CosmoPMC can be used as stand-alone program, it has the name nicaea (`http://www2.iap.fr/users/kilbinge/nicaea`).

Warning: Use undocumented features of the code at your own risk!

## 2. Installing CosmoPMC

### 2.1. Software requirements

CosmoPMC has been developed on GNU/Linux and Darwin/FreeBSD systems and should run on those architectures. Required are:

- C-compiler (e.g. `gcc`, `icc`)
- pmclib (Sect. 2.2)
- GSL (`http://www.gnu.org/software/gsl`), version 1.15 or higher
- FFTW (`http://www.fftw.org`)
- Message Parsing Interface (MPI) (`http://www-unix.mcs.anl.gov/mpi`) for parallel calculations

Optional:

- `csh`, for post-processing, auxiliary scripts; recommended
- `perl` (`http://www.perl.org`), for post-processing, auxiliary scripts; recommended
- `yorick` (`http://yorick.sourceforge.net`), post-processing, mainly plotting
- `python` (`http://www.python.org`), for running the configuration script
- `R` (`http://www.r-project.org`), post-processing

To produce 1D and 2D marginal posterior plots with scripts that come with CosmoPMC, either `yorick` or `R` are required.

Necessary for CMB anisotropies support:

- Fortran compiler (e.g. `ifort`)
- Intel Math Kernel libraries (`http://software.intel.com/en-us/intel-mkl`)
- CAMB (`http://camb.info,http://cosmologist.info/cosmomc`)
- WMAP data and likelihood code (`http://lambda.gsfc.nasa.gov`)

## 2.2. Download and install pmclib

The package pmclib can be downloaded from the CosmoPMC site `http://www.cosmopmc.info`.

After downloading, unpack the gzipped tar archive

```
> tar xzf pmclib_x.y.tar.gz
```

This creates the pmclib root directory `pmclib_x.y`. pmclib uses `waf`[1] instead of configure/make to compile and build the software. Change to that directory and type

```
> ./waf --local configure
```

See `./waf --help` for options. The packages `lua`, `hdf5` and `lapack` are optionally linked with pmclib but are not necessary to run CosmoPMC. Corresponding warnings of missing files can be ignored. Instead of a local installation (indicated by `--local`), a install prefix can be specified with `--prefix=PREFIX` (default `/usr/local`).

[1] `http://code.google.com/p/waf`

## 2.3. Patch pmclib

For CosmoPMC v1.2 and pmclib v1.x, a patch of the latter is necessary. From http://www.cosmopmc.info , download patch_pmclib_1.x_1.2.tar.gz and follow the instructions in the readme file readme_patch_pmclib_1.x_1.2.txt.

## 2.4. Download and install CosmoPMC

The newest version of CosmoPMC can be downloaded from the site http://www.cosmopmc.info.

First, unpack the gzipped tar archive

```
> tar xzf CosmoPMC_v1.2.tar.gz
```

This creates the the CosmoPMC root directory CosmoPMC_v1.2. Change to that directory and run

```
> [python] ./configure.py
```

This (poor man's) configure script copies the file Makefile.no_host to Makefile.host and sets host-specific variables and flags as given by the command-line arguments. For a complete list, see 'configure.py --help'.

Alternatively, you can copy by hand the file Makefile.no_host to Makefile.host and edit it. If the flags in this file are not sufficient to successfully compile the code, you can add more flags by rerunning configure.py, or by manually editing Makefile.main. Note that a flag in Makefile.main is overwritten if the same flag is present in Makefile.host.

To compile the code, run

```
> make; make clean
```

On success, symbolic links to the binary executables (in ./exec) will be set in ./bin.

It is convenient to define the environment variable COSMOPMC and to set it to the main CosmoPMC directory. For example, in the C-shell:

```
> setenv COSMOPMC /path/to/CosmoPMC_v1.2
```

This command can be placed into the startup file (e.g. ~/.cshrc for the C-shell). One can also add $COSMOPMC/bin to the PATH environment variable.

# 3. Running CosmoPMC

## 3.1. Quick reference guide

### Examples

To get familiar with CosmoPMC, use the examples which are contained in the package. Simply change to one of the subdirectories in `$COSMOPMC/Demo/MC_Demo` and proceed on to the point **Run** below.

### User-defined runs

To run different likelihood combinations, or your own data, the following two steps are necessary to set up a CosmoPMC run.

1. **Data and parameter files**

   Create new directory with `newdir_pmc.sh`. When asked, enter the likelihood/data type. More than one type can be chosen by adding the corresponding (bit-coded) type id's. Symbolic links to corresponding files in `$COSMOPMC/data` are set, and parameter files from `$COSMOPMC/par_files` are copied to the new directory on request.

   If necessary, copy different or additional data and/or parameter files to the present directory.

2. **Configuration file**

   Create the PMC configuration file `config_pmc`. Examples for existing data modules can be found in `$COSMOPMC/Demo/MC_Demo`, see also Sect. 5 for details.

   In some cases, information about the galaxy redshift distribution(s) have to be provided, and the corresponding files copied (see `$COSMOPMC/Demo` for example files '`nofz*`').

### Run

Type

```
> $COSMOPMC/bin/cosmo_pmc.pl -n NCPU
```

to run CosmoPMC on `NCPU` CPUs. See '`cosmo_pmc.pl -h`' for more options. Depending on the type of initial proposal (Sect. 3.2), a maximum-search is started followed by a Fisher matrix calculation. After that, PMC is started. Fig. 1 shows a flow chart of the script's actions.

### Diagnostics

Check the files `perplexity` and `enc`. If the perplexity reaches values of 0.8 or larger, and if the effective number of components (ENC) is not smaller than 1.5, the posterior has very likely been explored sufficiently. Those and other files are updated during run-time and can be monitored while PMC is running. See Sect. 3.3.1 for more details.

### Results

The text file `iter_{niter-1}/mean` contains mean and confidence levels. The file `iter_{niter-1}/all_contour2d.pdf` shows the 1d- and 2d-marginals. Plots can be redone or refined, or created from other than the last iteration with `plot_contour2d.pl`. Note that in the default setting, the posterior plots are not smoothed. See Sect. 6.1.1for more details, and for information on the alternative script `plot_confidence.R`.

## 3.2. CosmoPMC in detail

This section describes in more detail how PMC is run, and which decisions the user has to make before starting and after stopping a PMC run.

**Initial proposal** The choice of the initial proposal, used during the first PMC iteration, is of great importance for a successful PMC run. The following options are implemented, determined by the key '`sinitial`' in the configuration file (see Sect. 5):

1. **`sinitial = fisher_rshift`** The Fisher matrix is used as the covariance of a multi-variate Gaussian/Student-$t$ distribution $g$. A mixture-model is constructed by creating $D$ copies of $g$. Each copy is displaced from the ML point by a random uniform shift, and its variance is stretched by random uniform factor.
2. **`sinitial = fisher_eigen`** A mixture-model is constructed in a similar way as the first case, with the difference that the shift from the ML point is now performed along the major axes of the Fisher ellipsoid. Note that if the Fisher matrix is diagonal, the shift of each component only concerns one parameter.
3. **`sinitial = file`** The initial proposal is read from a file (of `mix_mvdens` format), e.g. from a previous PMC run.
4. **`sinitial = random_pos`** Mixture-model components with random variance (up to half the box size) and random positions. This case should only be used if the posterior is suspected to be multi-modal, or the calculation of the Fisher matrix fails.

In many cases, a mixture of multi-variate Gaussians as the proposal is the best choice. For that, set the degrees-of-freedom ($\nu$) parameter `df` to -1. For a posterior with heavy tails, a Student-$t$ distribution might be more suited. The degrees of freedom $\nu$ can be chosen freely; $\nu = 3$ is a common choice. For $\nu \to \infty$, a Gaussian distribution is reached asymptotically.

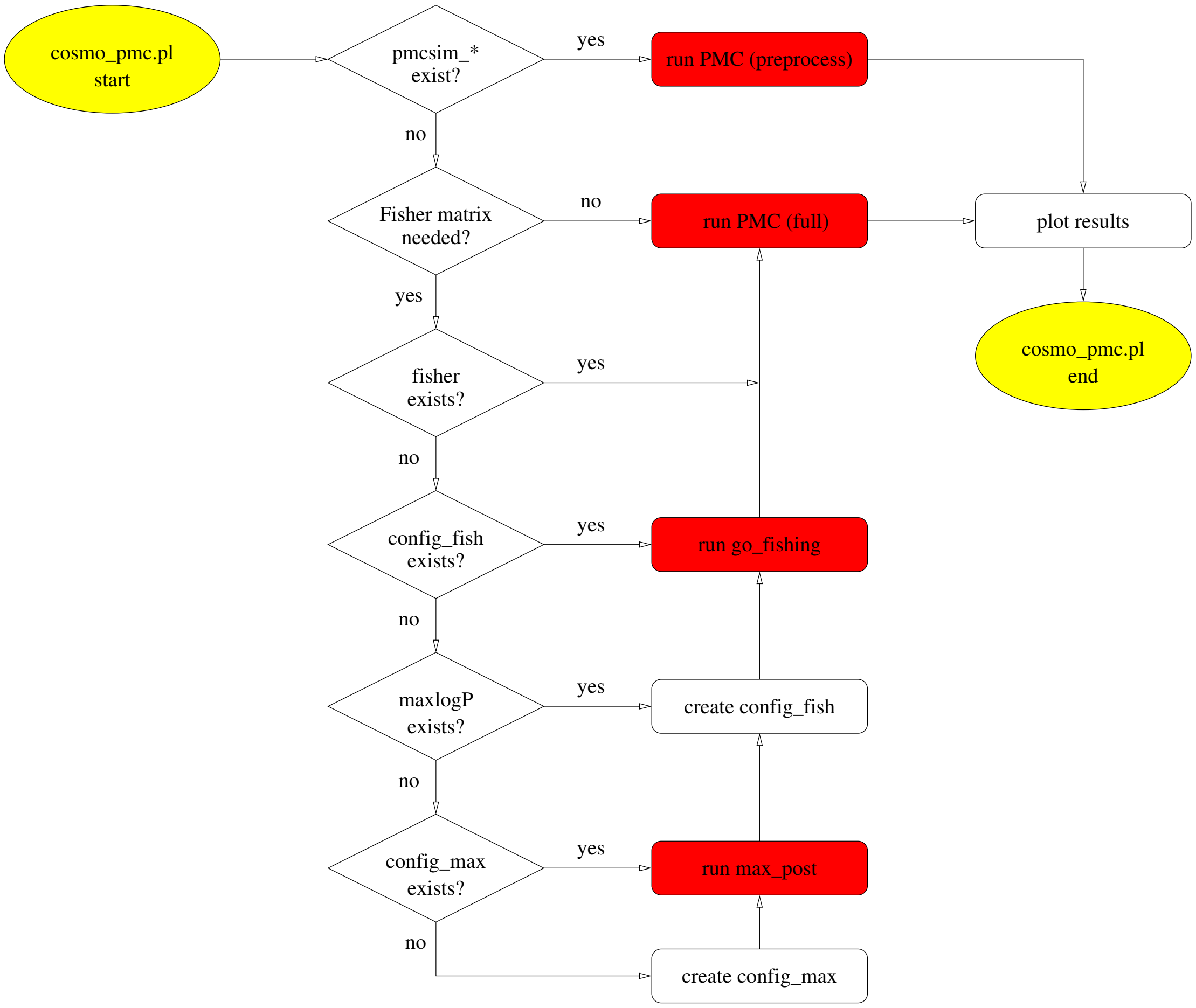

Figure 1: Flow chart for `cosmo_pmc.pl`.

If the Fisher matrix has to be calculated for the initial proposal, the script `cosmo_pmc.pl` calls `max_post` and `go_fishing` to estimate the maximum-likelihood point and the Hessian at that point, respectively. The script `config_pmc_to_max_and_fish.pl` can be used to create the corresponding configuration files from the PMC config file for manual calls of `max_post` and `go_fishing`.

**Updating the proposal** The PMC algorithm automatically updates the proposal after each iteration, no user interference is necessary.

The method to update the proposal is a variant of the Expectation-Maximization algorithm (EM, Dempster et al. 1977). It leads to an increase of the perplexity and an increase of ESS. Detailed descriptions of this algorithm in the case of multi-variate Gaussian and Student-*t* distributions can be found in Cappé et al. (2008) and Wraith et al. (2009).

**Dead components** A component can 'die' during the updating if the number of points sampled from that component is less than `MINCOUNT = 20`, or its weight is smaller than the inverse total number of sample points $1/N$. There are two possibilities to proceed. First, the component is 'buried', its weight set to zero so that no points are sampled from it in subsequent iterations. Alternatively, the component can be revived. In this case, it is placed near the component $\phi_{d_0}$ which has maximum weight, and it is given the same covariance as $\phi_{d_0}$.

The first case is the standard method used in Wraith et al. (2009). The second method tries to cure cases where the majority of components die. This can happen if they start too far off from the high-density posterior region. Often, only one component remains to the end, not capable of sampling the posterior reliably.

Both options can be chosen using the config file (Sect. 5) key `sdead_comp = {bury|revive}`.

**Errors** If an error occurs during the calculation of the likelihood, the error is intercepted and the likelihood is set to zero. Thus, the parameter vector for which the error occurs is attributed a zero importance weight and does not contribute to the final sample. An error message is printed to `stderr` (unless CosmoPMC is run with the option `-q`) and PMC continues with the next point.

An error can be due to cosmological reasons, e.g. a redshift is probed which is larger than the maximum redshift in a loitering Universe. Further, a parameter could be outside the range of a fitting formulae, e.g. a very small scalar spectral index in the dark matter transfer function.

Usually, the errors printed to `stderr` during PMC sampling can be ignored.

**Random numbers** The GSL random number generator is used to generate random variables. It is initialised with a seed reading the current time, to produce different (pseudo-) random numbers at each call. The seed is written to the log file. Using the option '`-s SEED`', a user-specified seed can be defined. This is helpful if a run is to be repeated with identical results.

## 3.3. Output files

Each iteration *i* produces a number of output files which are stored in subdirectories `iter_i` of the CosmoPMC starting directory. Files which are not specific to a single iteration are placed in the starting directory.

### 3.3.1. Diagnostics

Unlike in MCMC, with adaptive importance sampling one does not have to worry about convergence. In principle, the updating process can be stopped at any time. There are however diagnostics to indicate the quality and effectiveness of the sampling.

#### Perplexity and effective sample size `perplexity`

The perplexity $p$ is defined in eq. (18) of Wraith et al. (2009). The range of $p$ is $[0; 1]$, and will approach unity if the proposal and posterior distribution are close together, as measured by the Kullback-Leibler divergence. The initial perplexity is typically very low ($< 0.1$) and should increase from iteration to iteration. Final values of 0.99 and larger are not uncommon, but also for $p$ of about 0.6-0.8 very accurate results can be obtained. If $p$ is smaller than say 0.1, the PMC sample is most likely not representative of the posterior. Intermediate values for $p$ are not straight-forward to interpret.

Closely related to the perplexity is the effective sample size ESS, which lies in the range $[1; N]$. It is interpreted as the number of sample point with zero weight (Liu & Chen 1995). A large perplexity is usually accompanied by a high ESS. For a successful PMC run, ESS is much higher than the acceptance rate of a Monte Carlo Markov chain, which is typically between 0.15 and 0.25.

The file `perplexity` contains the iteration $i$, perplexity $p$, ESS for that iteration, and the total ESS. This file is updated after each iteration and can therefore be used to monitor a PMC run.

If there are points with very large weights, they can dominate the other points whose normalised weights will be small. Even a few sample points might dominate the sum over weights and result in a low perplexity. The perplexity is the most sensitive quantity to those high-weight points, much more than e.g. the mean, the confidence intervals or the evidence.

#### Effective number of proposal components `enc`

The proposal $q^t$ provides useful information about the performance of a PMC run. For example, the effective number of components, defined in complete analogy to ESS,

$$\mathrm{ENC} = \left( \sum_{d=1}^{D} \left\{ \alpha_d^t \right\}^2 \right)^{-1}, \tag{5}$$

is an indication of components with non-zero weight. If ENC is close to unity, the number of remaining components to sample the posterior is likely to be too small to provide a representative sample. For a badly chosen initial proposal, this usually happens already at the first few iterations. By monitoring the file `enc` which is updated each iteration, an unsuccessful PMC run can be aborted.

The effective number of components can also be determined from any proposal file (`mix_mvdens` format) with the script `neff_proposal.pl`.

An additional diagnostic is the evolution of the proposal components with iteration. This illustrates whether the components spread out nicely across the high-posterior region and reach a more or less stationary behaviour, or whether they stay too concentrated at one point. The scripts `proposal_mean.pl` (`proposal_var.pl`) read in the proposal information $q^t$ and plot the means (variances) as function of iteration $t$.

### 3.3.2. Results

#### PMC samples `iter_i/pmcsim`

This file contains the sample points. The first column is the (unnormalised) importance weight (log), the second column denotes the component number from which the corresponding point was sampled. Note that the $n_{\rm clip}$ points with highest weights are not considered in subsequent calculations (of moments, perplexity, evidence etc.). The next $p$ columns are the $p$-dimensional parameter vector. Optionally, $n_{\rm ded}$ numbers of deduced parameters follow.

#### Proposals `iter_i/proposal`

The proposal used for the importance sampling in iteration $i$ is in `mix_mvdens` format (Sect. A.3). The final proposal, updated from the sample of the last iteration, is `proposal_fin`.

#### Mean and confidence intervals `iter_i/mean`

This file contains mean and one-dimensional, left- and right-sided confidence levels (c.l.). A c.l. of $p\%$ is calculated by integrating the one dimensional normalised marginal posterior starting from the mean in positive or negative direction, until a density of $p\%/2$ is reached. PMC outputs c.l.'s for $p = 63.27\%, 95.45\%$ and $99.73\%$. With the program `cl_one_sided`, one-sided c.l.'s can be obtained.

For post-processing, the program `meanvar_sample` outputs the same information (mean and c.l.) from an existing PMC sample, including possible deduced parameters.

#### Resampled PMC simulations `iter_{niter-1}/sample`

If `cosmo_pmc.pl` has been run with the option `-p`, the directory of the final iteration contains the file of parameter vectors `sample`, which is resampled from the PMC simulation `pmcsim`, taking into account the importance weights. The resampled points all have unit weight. Resampling is a post-processing steop, it is performed by calling the `R` script `sample_from_pmcsimu.R` from `cosmo_pmc.pl`; this can also be done manually with any `pmscim` simulation.

### Histograms `iter_i/chi_j, iter_i/chi_j_k`

One- and two-dimensional histograms are written at each iteration to the text files `chi_j` and `chi_j_k`, respectively, where $j$ and $k$, $j < k$, are parameter indices. Those histograms can be used to create 1d- and 2d-marginals, using the script `plot_contour2d.pl`. The bin number is set by the config entry `nbinhist`.

In post-processing, use `histograms_sample` to produce histograms from a PMC sample. This can be useful if deduced parameters have been added to the sample.

### Covariance `iter_i/covar*.fin`

The parameter covariance and inverse covariance are printed to the files `covar.fin` and, respectively, `covarinv.fin`. The addition "`+ded`" in the file name indicates the inclusion of deduced parameters. The covariance matrices are in "mvdens"-format (see Sect. A.3).

### Evidence `evidence`

This file contains the Bayesian evidence as a function of iteration. Before the first iteration, the Laplace approximation using the Fisher matrix is printed to `evidence_fisher` if the file `fisher` exists. At each iteration $i$, `iter_i/evidence_covarinv` contains the Laplace approximation of the evidence from the inverse covariance matrix of the sample `iter_i/pmcsim`.

### 3.3.3. Deduced parameters

Deduced parameters can be part of a PMC simulation. These parameters are not sampling parameters, but they are deduced from the main parameters. For example, if $\Omega_\mathrm{m}$ and $\Omega_\Lambda$ are sampling parameters of a non-flat model, the curvature $\Omega_K = \Omega_\mathrm{m} + \Omega_\Lambda$ can be a deduced parameter.

In most cases, deduced parameters are ignored while running CosmoPMC. They are usually added to the PMC simulation after the sampling, for example using a script. In the case of galaxy clustering, `add_deduced_halomodel` adds deduced parameters which depend on the sampling parameters but also on the underlying cosmology and halo model.

A PMC simulation with deduced parameters added can be used as input to `histograms_sample`, to create the histogram files, now including the deduced parameters. These can then in turn be read by and `plot_contour2d.pl` to produce 1d- and 2d-marginals, including the deduced parameters. Alternatively, the PMC simulation with added parameters can be resampled using `sample_from_pmcsimu.R`, from which plots can be created by `plot_confidence.R`.

#### 3.3.4. Other files

**Maximum-posterior parameter** `max_logP`

`max_post` stores its estimate of the maximum posterior in this file.

**Fisher matrix** `fisher`

The final result of `go_fishing`, the Fisher matrix in mvdens (Sect. A.3) format.

**Log files** `log_max_post, log_fish, log_pmc`

`max_post`, `go_fishing` and `cosmo_pmc` each produce their corresponding log file.

# 4. Cosmology

The cosmology part of CosmoPMC is essentially the same as the stand-alone package NICAEA[2]. This excludes the external program `camb` and the WMAP likelihood library, which are called by CosmoPMC for CMB anisotropies. Further, CosmoPMC contains a wrapper layer to communicate between the PMC sampling and the cosmology modules.

## 4.1. Basic calculations

A number of routines to calculate cosmological quantities are included in the code. These are

- Background cosmology: Hubble parameter, distances, geometry
- Linear perturbations: growth factor, transfer function, cluster mass function, linear 3D power spectra
- Non-linear evolution: fitting formulae for non-linear power spectra (Peacock & Dodds 1996; Smith et al. 2003), emulators (Heitmann et al. 2009, 2010; Lawrence et al. 2010), halo model
- Galaxy clustering: HOD model
- Cosmic shear: convergence power spectrum, second-order correlation functions and derived second-order quantities, third-order aperture mass skewness
- CMB anisotropies via `camb`.

[2] `http://www2.iap.fr/users/kilbinge/nicaea`

Table 1: Extrapolation of the power spectra

| snonlinear | $k_{max}$ | $n_{ext}$ |
|---|---|---|
| linear | $333.6\,h$ Mpc$^{-1}$ | $n_s - 4$ |
| pd96 | $333.6\,h$ Mpc$^{-1}$ | $-2.5$ |
| smith03, smith03_de | $333.6\,h$ Mpc$^{-1}$ | Eq. (61), Smith et al. (2003) |
| coyote10 | 2.416 Mpc$^{-1}$ | no extrapolation |

#### 4.1.1. Density parameters

Both the density parameters ($\Omega_X = \rho_X/\rho_c$) and the physical density parameters ($\omega_x = \Omega_x h^2$) are valid input parameters for sampling with PMC. Internally, the code uses non-physical density parameters ($\Omega_X$). All following rules hold equivalently for both classes of parameters. Note that physical and non-physical density parameters can not be mixed, e.g. $\Omega_c$ and $\omega_K$ on input causes the program to abort.

The parameter for massive neutrinos, $\Omega_{\nu,\text{mass}}$, is not contained in the matter density $\Omega_m = \Omega_c + \Omega_b$.

A parameter which is missing from the input list is assigned the default value, found in the corresponding cosmology parameter file (cosmo.par), unless there is an inconsistency with other input parameters. E.g., if $\Omega_{de}$ and $\Omega_K$ are input parameters, $\Omega_m$ is assigned the value $\Omega_m = 1 - \Omega_{de} - \Omega_K - \Omega_{\nu,\text{mass}}$, to keep the curvature consistent with $\Omega_K$.

A flat Universe is assumed, unless (a) both $\Omega_m$ and $\Omega_{de}$, or (b) $\Omega_K$ are given as input parameter.

#### 4.1.2. Matter power spectrum

Usually, models of the non-linear power spectrum have a limited validity range in $k$ and/or redshift. For small $k$, each model falls back to the linear power spectrum, which goes as $P_\delta(k) \propto k^{n_s}$. For large $k$, the extrapolation as a power law $P_\delta(k) \propto n_{ext}$ is indicated in Table 4.1.2.

See for more details on the models.

**The Coyote emulator** In the coyote10 case, the power spectrum is zero for $k > k_{max}$. The same is true for redshifts larger than the maximum of $z_{max} = 1$. See Eifler (2011) for an alternative approach.

The Hubble constant $h$ can not be treated as a free parameter. For a given cosmology, it has to be fixed to match the CMB first-peak constraint $\ell_A = \pi d_{ls}/r_s = 302.4$, where $d_{ls}$ is the distance to last scattering, and $r_s$ is the sound horizon. This can be done with the function set_H0_Coyote, see Demo/lensingdemo.c for an example. When doing sampling with non-physical density parameters, $h$ has to be set at each sample point. Alternatively, the physical density parameters can be sampled, where $h$ is set internally to match the CMB peak.

### 4.1.3. Likelihood

Each cosmological probe has its own log-likelihood function. The log-likelihood function is called from a wrapping routine, which is the interface to the PMC sampler. In general, within this function the model vector is computed using the corresponding cosmology routine. The exception are the WMAP-modules where the $C_\ell$'s are calculated using `camb` and handed over to the log-likelihood function as input.

## 4.2. Cosmic shear

CosmoPMC implements second- and third-order weak lensing observables.

### 4.2.1. Second-order

The basic second-order quantities in real space for weak gravitational lensing are the two-point correlation functions $\xi_\pm$ (2PCF) (e.g Kaiser 1992),

$$\xi_\pm(\theta) = \frac{1}{2\pi} \int_0^\infty \mathrm{d}\ell \, \ell P_\kappa(\ell) \mathrm{J}_{0,4}(\ell\theta). \tag{6}$$

Data corresponding to both functions (`slensdata=xipm`) as well as only one of them (`xip`, `xim`) can be used. The aperture-mass dispersion (Schneider et al. 1998)

$$\langle M_\mathrm{ap}^2 \rangle(\theta) = \frac{1}{2\pi} \int_0^\infty \mathrm{d}\ell \, \ell P_\kappa(\ell) \hat{U}^2(\theta\ell) \tag{7}$$

is supported for two filter functions $U_\theta(\vartheta) = u(\vartheta/\theta)/\theta^2$ (Schneider et al. 1998; Crittenden et al. 2002),

$$\text{polynomial (map2poly):} \quad u(x) = \frac{9}{\pi}(1 - x^2)\left(\frac{1}{3} - x^2\right)\mathrm{H}(1 - x); \tag{8}$$

$$\text{Gaussian (map2gauss):} \quad u(x) = \frac{1}{2\pi}\left(1 - \frac{x^2}{2}\right)\mathrm{e}^{-\frac{x^2}{2}}. \tag{9}$$

The top-hat shear dispersion (Kaiser 1992)

$$\langle|\gamma|^2\rangle_\mathrm{E,B}(\theta) = \frac{1}{2\pi} \int_0^\infty \mathrm{d}\ell \, \ell \, P_\kappa(\ell) \, \frac{4\mathrm{J}_1(\ell\theta)}{(\ell\theta)^2} \tag{10}$$

is used with `slensdata = gsqr`.

Pure E-/B-mode separating functions (Schneider & Kilbinger 2007) are chosen with `slensdata = decomp_eb`. For the lack of analytical expressions for filter functions to obtain these real-space statistics from the convergence power spectrum, they are calculated by integrating over the 2PCF. The integral is performed over the finite angular interval $[\vartheta_{\rm min}; \vartheta_{\rm max}]$. The prediction for the E-mode is

$$E = \frac{1}{2} \int_{\vartheta_{min}}^{\vartheta_{max}} \mathrm{d}\vartheta \, \vartheta \, \left[ T_+(\vartheta) \, \xi_+(\vartheta) \pm T_-(\vartheta) \, \xi_-(\vartheta) \right]. \qquad (11)$$

Two variants of filter functions are implemented: The 'optimized' E-/B-mode function Fu & Kilbinger (2010) for which the real-space filter functions are Chebyshev polynomials of the second kind,

$$T_+(\vartheta) = \tilde{T}_+ \left( x = \frac{2\vartheta - \vartheta_{\rm max} - \vartheta_{\rm min}}{\vartheta_{\rm max} - \vartheta_{\rm min}} \right) = \sum_{n=0}^{N-1} a_n U_n(x); \qquad U_n(x) = \frac{\sin[(n+1) \arccos x]}{\sin(\arccos x)}. \qquad (12)$$

The coefficients $a_n$ have been optimized with respect to signal-to-noise and the $\Omega_{\rm m}$-$\sigma_8$ Fisher matrix. The function $E$ is defined as a function of the lower angular limit $\vartheta_{\rm min}$. The ratio $\eta$ of lower to upper limit, $\eta = \vartheta_{\rm min}/\vartheta_{\rm max}$ is fixed.

The second variant are the so-called COSEBIs (Complete Orthogonal Sets of E-/B-mode Integrals; Schneider et al. 2010). We implement their 'logarithmic' filter functions,

$$T^{\rm log}_{+,n}(\vartheta) = t^{\rm log}_{+,n} \left[ z = \ln \left( \frac{\vartheta}{\vartheta_{\rm min}} \right) \right] = N_n \sum_{j=0}^{n+1} c_{nj} z^j = N_n \prod_{j=1}^{n+1} (z - r_{nj}).. \qquad (13)$$

The coefficients $c_{nj}$ are fixed by integral conditions that assure the E-/B-mode decomposition of the 2PCF on a finite angular integral. They are given by a linear system of equations, which is given in Schneider et al. (2010). To solve this system, a very high numerical accuracy is needed. The Mathematica notebook file `$COSMOPMC/par_files/COSEBIs/cosebi.nb`, adapted from Schneider et al. (2010), can be run to obtain the coefficients for a given $\vartheta_{\rm min}$ and $\vartheta_{\rm max}$. An output text file is created with the zeros $r_{ni}$ and amplitudes $N_n$. The file name is `cosebi_tplog_rN_[Nmax]_[thmin]_[thmax]`, where `Nmax` is the number of COSEBI modes, `thmin` and `thmax` are the minimum and maximum angular scale $\vartheta_{\rm min}$ and $\vartheta_{\rm max}$, respectively. For a given $\vartheta_{\rm min}$ and $\vartheta_{\rm max}$, specified with the config entries `th_min` and `th_max`, CosmoPMC reads the corresponding text file from a directory that is specified by `path`. A sample of files with various scales are provided in `$COSMOPMC/par_files/COSEBIs`.

The COSEBIs are discrete numbers, they are specified by an integer mode number $n$.

In both cases of pure E-/B-mode separating statistics, the function $T_-$ is calculated from $T_+$ according to Schneider et al. (2002).

The additional flag `decomp_eb_filter` decides between different filter functions:

| decomp_eb_filter | Reference | Filter function typ | $\eta$ |
|---|---|---|---|
| FK10_SN | Fu & Kilbinger (2010) | optimized Signal-to-noise | 1/50 |
| FK10_FoM_eta10 | Fu & Kilbinger (2010) | optimized Fisher matrix | 1/10 |
| FK10_FoM_eta50 | Fu & Kilbinger (2010) | optimized Fisher matrix | 1/50 |
| COSEBIs_log | Schneider et al. (2010) | logarithmic | |

The convergence power spectrum $P_\kappa$ with covariance matrix can be used with the flag `slensdata = pkappa`.

#### 4.2.2. Third-order

We implement the aperture-mass skewness (Pen et al. 2003; Jarvis et al. 2004; Schneider et al. 2005) with the Gaussian filter (eq. **??**). There are two cases:

- `slensdata = map3gauss`
  The 'generalised' skewness $\langle M_{\rm ap}^3 \rangle (\theta_1, \theta_2, \theta_3)$ (Schneider et al. 2005) with three filter scales.
- `slensdata = map3gauss_diag`
  The 'diagonal' skewness $\langle M_{\rm ap}^3 \rangle (\theta)$ using a single aperture filter scale.

**TODO: equations**

#### 4.2.3. Second- plus third-order

A joint data vector of second- and third-order observables can be used in CosmoPMC. The covariance is interpreted as a joint block matrix, with the second-order and third-order auto-covariances on the diagonal, and the cross-correlation on the off-diagonal blocks. The possible scenarios are:

- `slensdata = map2gauss_map3gauss`
  Gaussian aperture-mass dispersion and generalised skewness.
- `slensdata = map2gauss_map3gauss_diag`
  Gaussian aperture-mass dispersion and diagonal skewness.
- `slensdata = decomp_eb_map3gauss`
  Log-COSEBIs and generalised aperture-mass skewness. The flag `decomp_eb_filter` has to be set to `COSEBIs_log`.
- `slensdata = decomp_eb_map3gauss_diag`
  Log-COSEBIs and diagonal aperture-mass skewness. The flag `decomp_eb_filter` has to be set to `COSEBIs_log`.

The first two cases use the same filter for second- and third-order, and provide therefore a consistent measure for both orders. The last two cases use the optimal E-/B-mode function known for second order.

#### 4.2.4. Covariance

The covariance matrix is read from a file, and the inverse is calculated in CosmoPMC. The matrix has to be positive definite. An Anderson-Hartlap debiasing factor is multiplied to the inverse (Anderson 2003; Hartlap et al. 2007), which is specified with the config entry `corr_invcov`.This can also be used to rescale the covariance, e.g. to take into account a different survey area. Set this value to unity if no correction is desired.

The covariance is either taken to be constant and not dependent on cosmology. In that case, set `scov_scaling` to `cov_const`. Or the approximated schemes from Eifler et al. (2009) are adopted, see **?** for the implementation. In that scheme, the shot-noise term $D$ is constant, the mixed term $M$ is modulated with $\Omega_\mathrm{m}$ and $\sigma_8$ using fitting formluae, and the cosmic-variance term $V$ is proportional to the square of the shear correlation function. This scheme is available for `slensdata = xipm`. The three covariance terms have to be read individually. The entry `covname`, which for `scov_scaling = cov_const` corresponds to the total covariance matrix, now specified the file name of cosmic-variance term, `covname_M` the name of the mixed term, and `covname_D` the name of the shot-noise term.

#### 4.2.5. Reduced shear

The fact that not the shear $\gamma$ but the reduced shear $g = \gamma/(1-\kappa)$ is observable leads to corrections to the shear power spectrum of a few percent, mainly on small scales. These corrections are either ignored, or modelled to first order according to Kilbinger (2010). This is controlled in the lensing parameter file (`cosmo_lens.par`). The parameter range where the reduced-shear corrections are valid are indicated in Table 2.

#### 4.2.6. Angular scales

The flag `sformat` describes the mapping of angular scales (given in the data file) and 'effective' scales, where the model predictions of the shear functions are evaluated:

1. `sformat = angle_center`: The effective scale is the same as given in the data file, $\theta_\mathrm{eff} = \theta$.
2. `sformat = angle_mean`: The model is averaged over a range of scales $[\theta_0, \theta_1]$ given in the data file.

Table 2: Parameter limits where the reduced-shear corrections are valid (from Kilbinger 2010).

| $\alpha$ | Parameter | lower | upper |
|---|---|---|---|
| 1 | $\Omega_{\rm m}$ | 0.22 | 0.35 |
| 2 | $\Omega_{\rm de}$ | 0.33 | 1.03 |
| 3 | $w$ | -1.6 | -0.6 |
| 4 | $\Omega_{\rm b}$ | 0.005 | 0.085 |
| 5 | $h$ | 0.61 | 1.11 |
| 6 | $\sigma_8$ | 0.65 | 0.93 |
| 7 | $n_{\rm s}$ | 0.86 | 1.16 |

3. `sformat = angle_wlinear`: The model is the weighted average over a range of scales $[\theta_0, \theta_1]$, where the weight is $w = \theta/\mathrm{arcmin}$.
4. `sformat = angle_wquadr`: The model is the weighted average over a range of scales $[\theta_0, \theta_1]$, where the weight is $w = a_1(\theta/\mathrm{arcmin}) + a_2(\theta/\mathrm{arcmin})^2$.

The first mode (`angle_center`) should be used for aperture-mass, shear rms and 'ring' statistics, since those quantities are not binned, but instead are integrals up to some angular scale $\theta$. For the correlation functions, in particular for wide angular bins, one of the last three modes is preferred. The quadratic weighting (`angle_wquadr`) corresponds to a weighting of the correlation function by the number of pairs[3]. This mode was used in the COSMOS analysis (Schrabback et al. 2010).

## 4.3. SNIa

The standard distance modulus (`schi2mode = chi2_simple`) for a supernova with index $i$ is

$$\mu_{B,i} = m^*_{B,i} - \bar{M} + \alpha(s_i - 1) - \beta c_i. \tag{14}$$

where the quantities measured from the light-curve fit are the rest-frame $B$-band magnitude $m^*_{B,i}$, the shape or stretch parameter $s_i$, and the color $c_i$. The universal absolute SNIa magnitude is $\bar{M}$, the linear response parameters to stretch and color are $\alpha$ and $\beta$, respectively. The $\chi^2$-function is

$$\chi^2_{\rm sn}(\boldsymbol{p}) = \sum_i \frac{\left[\mu_{B,i}(\boldsymbol{p}) - 5\log_{10}\left(\frac{d_{\rm L}(z_i,\boldsymbol{p})}{10\,\mathrm{pc}}\right)\right]^2}{\sigma^2(\mu_{B,i}) + \sigma^2_{{\rm pv},i} + \sigma^2_{\rm int}}, \tag{15}$$

where $d_{\rm L}$ is the luminosity distance and $z_i$ the redshift of object $i$. The contributions to the total error for object $i$ are: (1) The light-curve parameter variance $\sigma^2(\mu_{B,i}) = \boldsymbol{\theta}_2^{\rm t} W_2 \boldsymbol{\theta}_2$ with the parameter vector $\boldsymbol{\theta}_2 = (1, \alpha, \beta)$ and the covariance $W_2$ of the data vector $(m^*_{B,i}, s_i, c_i)$. (2) The peculiar velocity uncertainty $\sigma_{\mathrm{pv},i} = 5/\ln 10 \cdot v_\mathrm{p}/(c\, z_i)$. (3) The intrinsic absolute magnitude scatter $\sigma_\mathrm{int}$.

[3] P. Simon, private communication

The Hubble parameter is absorbed into the absolute magnitude which we define as $M = \bar{M} - 5 \log_{10} h_{70}$.

The form of this log-likelihood function has been used in Astier et al. (2006).

The following variations of the distance modulus and log-likelihood are implemented:

- `schi2mode = chi2_Theta1`: The $\chi^2$ is extended to include photometric zero-point uncertainties, see Kilbinger et al. (2009).
- `schi2mode = chi2_Theta2_denom_fixed`: The parameters $\alpha$ and $\beta$ in the denominator of (15) are fixed and kept constant during the Monte-Carlo sampling.
- `schi2mode = chi2_no_sc`: The stretch and color parameters are ignored, the distance modulus is $\mu_{B,i} = m^*_{B,i} - \bar{M}$.
- `schi2mode = chi2_betaz`: Instead of a single parameter, the color response is redshift-dependent, $\beta \to \beta + \beta_z z_i$.
- `chi2_dust`: Intergalactic dust absorption is taken into account in the distance modulus, see Ménard et al. (2010).

The covariance matrix $W_2$ of the data vector $(m^*_{B,i}, s_i, c_i)$ depends on the parameters $\alpha$ and $\beta$. In a Bayesian framework, this leads to an additional term $\frac{1}{2} \log \det W_2$ in the log-likelihood function. Taking into account this parameter-dependent term leads however to a biased maximum-likelihood estimator, in particular for $\alpha$ and $\beta$[4]. Therefore, it is recommended to not include this term. Use the flag `add_logdetCov = 0/1` in the configuration file to disable/enable this term.

## 4.4. CMB anisotropies

The full CMB anisotropies are handled externally: The $C_\ell$'s are calculated by calling `camb`[5] (Lewis et al. 2000), the WMAP likelihood function (3rd-, 5th- and 7th-year) is computed using the WMAP public code[6] (Dunkley et al. 2009). The maximum $\ell$ up to which the $C_\ell$'s are calculated and used in the likelihood can be determined in the configuration file. An $\ell_\mathrm{max} = 2000$ is recommended for high precision calculations.

The power spectrum from the Sunyaev-Zel'dovich (SZ) effect can be added to the $C_\ell$'s, multiplied with an amplitude $A$ as free parameter. The predicted SZ power spectrum is taken from Komatsu & Seljak (2002). This model has been used in the 3-, 5- and 7-year analyses of the WMAP data (Komatsu et al. 2011).

[4] J. Guy, private communication
[5] `http://camb.info`
[6] `http://lambda.gsfc.nasa.gov`

Alternatively, the WMAP distance priors (Komatsu et al. 2009) can be employed.

## 4.5. Galaxy clustering

### 4.5.1. Halomodel and HOD

The theoretical model of galaxy clustering is the one used in Coupon et al. (2012); see this paper for details of the model and further references.

As the basis to describe galaxy clustering, we implement the halo-model as reviewed in (Cooray & Sheth 2002), which accounts for the clustering of dark-matter halos. On top of that, a halo occupation distribution (HOD) function (Berlind & Weinberg 2002; Kravtsov et al. 2004; Zheng et al. 2005) is the prescription of how galaxies populate those halos. This function is the number of galaxies $N$ in a halo of mass $M$. With the flag `hod = berwein02_excl`, this number is expressed as the sum of central ($N_\mathrm{c}$) plus satellite ($N_\mathrm{s}$) galaxies,

$$N(M) = N_\mathrm{c}(M) \times [1 + N_\mathrm{s}(M)] \; , \tag{16}$$

with

$$n_\mathrm{c}(M) = \frac{1}{2}\left[1 + \mathrm{erf}\left(\frac{\log_{10} M - \log_{10} M_\mathrm{min}}{\sigma_{\log M}}\right)\right]; \tag{17}$$

$$n_\mathrm{s}(M) = \begin{cases} \left(\frac{M-M_0}{M_1}\right)^\alpha; & \text{if } M > M_0 \\ 0 & \text{else} \end{cases} \; , \tag{18}$$

We further compute the galaxy two-point correlation function $\xi(r)$ and its angular projection $w(\theta)$ using the redshift distribution provided by the user, as well as the galaxy number density (for a full description of the model see Coupon et al. 2012). To prevent haloes from overlapping, we implement the halo exclusion formalism as described in Tinker et al. (2005).

For the halo bias, three options are available:

- `shalo_bias = bias_sc`

  Bias expansion from the spherical collapse model, see e.g. eq. (68) from Cooray & Sheth (2002).

- `shalo_bias = bias_tinker05`

  Bias calibrated with numerical simulations, Tinker et al. (2005) eq. (A1).

- `shalo_bias = bias_tinker10`

  Updated bias fitting formua from Tinker et al. (2010), eq. (6) and Table 2.

The mass function describes the number of halos for a given mass and redshift. It is defined as

$$\frac{\mathrm{d}n}{\mathrm{d}\ln M} = \frac{\overline{\rho}_0}{M}\frac{\nu f(\nu)}{\nu}\frac{\mathrm{d}\nu}{\mathrm{d}\ln M}, \tag{19}$$

where $\nu(M, z) = \delta_c(z)/[D_+(z)\sigma(M)]$ is a measure of the overdensity with $\sigma(M)$ being the rms matter fluctuation in a top-hat window containing the mass $M$. $\overline{\rho}_0 = \Omega_m \rho_{c,0}$ is the mean density of matter at the present day.

The following mass functions are implemented, via the flag `smassfct`:

- From the spherical/eliptical collapse model:

$$\nu f(\nu) = A \sqrt{\frac{2}{\pi a \nu^2}} \left[1 + (a\nu^2)^{-p}\right] \exp\left(-\frac{a\nu^2}{2}\right), \tag{20}$$

  - `ps`: $p = 0, q = 1$ (Press & Schechter 1974)
  - `st`: $p = 0.3, q = 0.75$ (Sheth & Tormen 1999)
  - `st2`: $p = 0.3, q = 0.707$ (Sheth & Tormen 1999)

- From numerical simulations:

$$\nu f(\nu) = f(\sigma) = 0.315 \exp\left[-|\ln(\sigma^{-1} + 0.61|^{3.8}\right] \tag{21}$$

  - `j01`: (Jenkins et al. 2001)

The dark-matter halos have the density profile

$$\rho(r) = \rho_s \left[(r/r_s)^\alpha (1 + r/r_s)^{3-\alpha}\right]^{-1}. \tag{22}$$

For slopes unequal to the Navarro et al. (1997) value of $\alpha = 1$, closed expressions for the Fourier transform of $\rho$ do not exist, and the code will be slower.

The concentration parameter is given by

$$c(M, z) = \frac{c_0}{1 + z} \left[\frac{M}{M_\star}\right]^{-\beta}, \tag{23}$$

following Takada & Jain (2003). The parameters $c_0$ and $\beta$ can be chosen freely in the halomodel parameter file `halomodel.par`.

The log-likelihood function is the sum of the contribution from the angular correlation function and the galaxy number density $n_{gal}$:

$$\chi^2 = \sum_{i,j} \left[w^{obs}(\theta_i) - w^{model}(\theta_i)\right] \left(C^{-1}\right)_{ij} \left[w^{obs}(\theta_j) - w^{model}(\theta_j)\right] + \frac{\left[n_{gal}^{obs} - n_{gal}^{model}\right]^2}{\sigma_{n_{gal}}^2}, \tag{24}$$

where $n_{gal}^{model}$ is estimated at the mean redshift of the sample.

The number of galaxies (second term in eq. 24) can be included in the following way, with the config flag `sngal_fit_type`:

- `ngal_lin_fit`: linear (standard; according to the above equation)
- `ngal_log_fit`: logarithmical
- `ngal_no_fit`: no inclusion, second term is omitted
- `ngal_lin_fit_only`: exclusive, first term is omitted

### 4.5.2. Deduced parameters

The following deduced parameters can be computed:

- Mean galaxy bias
$$b_\mathrm{g}(z) = \int \mathrm{d}M\, b_\mathrm{h}(M,z)\, n(M,z) \frac{N(M)}{n_\mathrm{gal}(z)}, \tag{25}$$
where $b_\mathrm{h}$ is the halo bias, and
$$n_\mathrm{gal}(z) = \int N(M)\, n(M,z)\, \mathrm{d}M \tag{26}$$
is the total number of galaxies.
- Mean halo mass
$$\langle M_\mathrm{halo}\rangle(z) = \int \mathrm{d}M\, M\, n(M,z) \frac{N(M)}{n_\mathrm{gal}(z)}. \tag{27}$$
- Fraction of satellite galaxies
$$f_\mathrm{s}(z) = 1 - f_\mathrm{c}(z); \quad f_\mathrm{c}(z) = \int \mathrm{d}M\, n(M,z) \frac{N_\mathrm{c}(M)}{n_\mathrm{gal}(z)}. \tag{28}$$

Use the program `add_deduced_halomodel` to add those deduced parameters to a PMC sample. See the example config file `config_pmc_ded` in `Demo/MC_Demo/HOD/CFHTLS-T06`.

### 4.5.3. Clustering data

The angular two-point correlation function $w(\theta)$ is implemented, with the flag `shalodata = woftheta`. The measured (input) data $w_\mathrm{mes}$ is corrected for the integral constraint, via
$$w(\theta) = w_\mathrm{mes}(\theta) + w_C, \tag{29}$$
assuming that the measured correlation function can be fit by a power law
$$w_\mathrm{mes}(\theta) \approx A_w \left(\theta^{-\delta} - C\right). \tag{30}$$

The program `haloplot` outputs the correlation functions $w(\theta)$ and $\xi(r)$, the HOD function $N(M)$, and deduced parameters for given HOD input parameters.

### 4.5.4. Comoving volume

The comoving volume is needed to calculate the comoving number density of galaxies, following from the halomodel and the HOD parameters. There are two possibilities to calculate the comoving volume $V_C$. First, if $z_{min}$ and $z_{max}$ are larger than zero in the HOD parameter file `halomodel.par` (see Table 7), $V_C$ is computed between those two redshifts. Second, if both numbers are $< 0$, $V_C$ is weighted by the redshift distribution $n(z)$, see e.g. eq. (28) in Ross & Brunner (2009). In this weighting, the maximum value of $n(z)$ is set to unity.

## 4.6. BAO

BAO constraints are implemented with two distance measures:

- `smethod = distance_A`

  The distance parameter $A$ is defined in Eisenstein et al. (2005) as

  $$A(z) = \frac{D_V(z)}{c/H_0} \frac{\sqrt{\Omega_m}}{z} \tag{31}$$

  where

  $$D_V(z) = \left[ f_K^2[w(z)] \frac{cz}{H(z)} \right]^{1/3} \tag{32}$$

  is the spherically averaged distance to redshift $z$.

- `smethod = distance_d_z`

  The distance parameter $d$ is the ratio of sound horizon $r_s$ at drag epoch $z_d$ to spherically averaged distance (e.g. Percival et al. 2007),

  $$d(z) = \frac{r_s(z_d)}{D_V(z)}. \tag{33}$$

  We use the fitting formula for the drag redshift $z_d$ from Eisenstein & Hu (1998) and calculate the sound horizon as the distance a sound wave can travel prior to $z_d$ by numerical integration.

## 4.7. Redshift distribution

Some of the cosmology modules require a redshift distribution, for example lensing and HOD. Table 3 lists the implemented redshift distributions $n(z)$, via the flag `nofz`.

Each redshift bin can have a different type. The syntax for a redshift bin file is described in Appendix A.1.5.

Table 3: Redshift distribution types

| nofz | Description | $n(z) \propto \ldots$ | parameter list |
|---|---|---|---|
| hist | Histogram | $\sum_{i=0}^{n-1} N_i \cdot \mathbb{1}_{[z_i;z_{i+1}]}$ | (see text) |
| single | Single redshift | $\delta_{\rm D}(z - z_0)$ | $z_0, z_0$ |
| ludo | Fitting function | $(z/z_0)^\alpha \exp\left[-(z/z_0)^\beta\right]$ | $z_{\rm min}, z_{\rm max}, \alpha, \beta, z_0$ |
| jonben | | $z^a / \left(z^b + c\right)$ | $z_{\rm min}, z_{\rm max}, a, b, c$ |
| ymmk | | $\left(z^a + z^{ab}\right) / \left(z^b + c\right)$ | $z_{\rm min}, z_{\rm max}, a, b, c$ |

All redshift distributions are internally normalised as

$$\int_{z_{\rm min}}^{z_{\rm max}} {\rm d}z \, n(z) = 1. \tag{34}$$

### 4.8. CMB and the power spectrum normalisation parameter

The power spectrum normalisation parameter taken as input for CAMB is $\Delta^2_{\mathcal{R}}$, which is the amplitude of curvature perturbations at the pivot scale $k_0 = 0.002\,{\rm Mpc}^{-1}$. For lower-redshift probes such as lensing or HOD, the normalisation is described by $\sigma_8$, the rms fluctuation of matter in spheres of 8 Mpc/$h$. To combine those probes in a PMC run, $\Delta^2_{\mathcal{R}}$ has to be an input parameter, and $\sigma_8$ a deduced parameter. CMB has to come first in the list of data sets so that CAMB can calculate $\sigma_8$, which in turn is handed over to the lensing likelihood.

### 4.9. Parameter files

Tables 4 - 6 list the contents of the parameter files for basic cosmology, lensing, SNIa and HOD. Proto-types can be found in `$COSMOPMC/par_files`. These files specify the default values of parameters and flags. These default values are over-written if any of those parameter is used for Monte-Carlo sampling.

## 5. The configuration file

The programs `max_post`, `go_fishing`, `cosmo_pmc`, and `cosmo_mcmc` read a configuration file on startup. Each configuration file consist of two parts:

The first, basic part is common to all four config file types (Table 9). It consists of (1) the parameter section, (2) the data section and (3) the prior section. The data-specific entries in the data section are listed in Table 11.

Table 4: Basic cosmology parameter file (cosmo.par)

| | | |
|---|---|---|
| Omega_m | $\Omega_\mathrm{m}$ | Matter density, cold dark matter + baryons |
| Omega_de | $\Omega_\mathrm{de}$ | Dark-energy density (if $w = -1$, corresponds to $\Omega_\Lambda$ |
| w0_de | $w_0$ | Dark-energy equation-of-state parameter (constant term) |
| w1_de | $w_1$ | Dark-energy equation-of-state parameter (linear term, see sde_param) |
| h_100 | $h$ | Dimensionless Hubble parameter |
| Omega_b | $\Omega_\mathrm{b}$ | Baryon density |
| Omega_nu_mass | $\Omega_{\nu,\mathrm{mass}}$ | Massive-neutrino density (so far only for CMB) |
| N_eff_nu_mass | $N_{\mathrm{eff},\nu,\mathrm{mass}}$ | Effective number of massive neutrinos (so far only for CMB) |
| normalization | $\sigma_8$ | Power-spectrum normalisation at small scales (for normmode==0, see below) |
| n_spec | $n_\mathrm{s}$ | Scalar power-spectrum index |
| snonlinear | | Power spectrum prescription |
| | linear | Linear power spectrum |
| | pd96 | Peacock & Dodds (1996) |
| | smith03 | Smith et al. (2003) |
| | smith03_de | Smith et al. (2003) + dark-energy correction from icosmo.org |
| | coyote10 | 'Coyote Universe', Heitmann et al. (2009), Heitmann et al. (2010), Lawrence et al. (2010) |
| stransfer | | Transfer function |
| | bbks | Bardeen et al. (1986) |
| | eisenhu | Eisenstein & Hu (1998) 'shape fit' |
| | eisenhu_osc | Eisenstein & Hu (1998) with BAO wiggles |
| sgrowth | | Linear growth factor |
| | heath | Heath (1977) fitting formula |
| | growth_de | Numerical integration of differential equation for $\delta$ (recommended) |
| sde_param | | Dark-energy parameterisation |
| | jassal | $w(a) = w_0 + w_1 a(1 - a)$ |
| | linder | $w(a) = w_0 + w_1(1 - a)$ |
| normmode | | Normalization mode. 0: normalization=$\sigma_8$ |
| a_min | $a_\mathrm{min}$ | Minimum scale factor |

The second part is type-specific. See Table 10 for the PMC part, and Table 13 for the MCMC part. Example files can be found in subdirectories of $COSMOPMC/Demo/MC_DEMO.

Table 5: Weak lensing parameter file (cosmo_lens.par)

| | | |
|---|---|---|
| cosmo_file | | Basic cosmology file name (cosmo.par) |
| nofz_file | | Redshift distribution master file |
| redshift module[a] | | (see Table 8) |
| stomo | | Tomography correlations |
| | tomo_all | All correlations |
| | tomo_auto_only | Only auto-correlations (*ii*) |
| | tomo_cross_only | Only cross-correlations ($i \neq j$) |
| sreduced | | Reduced-shear treatment |
| | none | No correction |
| | K10 | Fitting-formulae from Kilbinger (2010) |
| q_mag_size[b] | $q$ | Magnification-bias coefficient, $q = 2(\alpha + \beta - 1)$ (see Kilbinger 2010, eq. 16) |

Table 6: SNIa parameter file (cosmo_SN.par)

| | | |
|---|---|---|
| cosmo_file | | Basic cosmology file name (cosmo.par) |
| Theta2 | $-M \; \alpha \; -\beta \; \beta_z$ | Distance modulus parameters |

To create a config file of type max_post or go_fishing from a PMC config file, the script config_pmc_to_max_and_fish.pl can be used.

Some flags are handled internally as integers (enumerations), but identified and set in the config file with strings. The corresponding key word carries the same name as the internal variable, preceded with an 's', e.g. the integer/string pair lensdata/slensdata.

The prior file, indicated if desired with the flag sprior, is a file in mvdens format. It specifies a Gaussian prior with mean and covariance as given in the file. Note that the covariance and not the inverse covariance is expected in the file.

Table 7: HOD parameter file (halomodel.par)

| | | |
|---|---|---|
| cosmo_file | | Basic cosmology file name (cosmo.par) |
| nofz_file | | Redshift distribution master file |
| redshift module[a] | | (see Table 8) |
| zmin | $z_{\rm min}$ | Minimum redshift (-1 if read from nzfile) |
| zmax | $z_{\rm max}$ | Maximum redshift (-1 if read from nzfile) |
| alpha_NFW | $\alpha$ | Halo density profile slope ($\alpha = 1$ for NFW) |
| c0 | $c_0$ | Concentration parameter at $z = 0$ |
| beta_NFW | $\beta$ | Concentration parameter slope of mass dependence |
| smassfct | | Halo mass function type |
| | ps | (Press & Schechter 1974), $p = 0, q = 1$ |
| | st | (Sheth & Tormen 1999), $p = 0.3, q = 0.75$ |
| | st2 | (Sheth & Tormen 1999), $p = 0.3, q = 0.707$ |
| | j01 | (Jenkins et al. 2001) |
| M_min | $M_{\rm min}$ | Minimal mass for central galaxies $[h^{-1} M_\odot]$ |
| M1 | $M_1$ | Scale mass for satellites $[h^{-1} M_\odot]$ |
| M0 | $M_0$ | Minimum mass for satellites $[h^{-1} M_\odot]$ |
| sigma_log_M | $\sigma_{\log M}$ | Logarithmic dispersion for central galaxies |
| alpha | $\alpha$ | Slope for satellite mass dependence |
| shod | | HOD type |
| | berwein02_hexcl | Berlind & Weinberg (2002) with halo exclusion |

Table 8: Redshift module file (nofz.par)

| | | |
|---|---|---|
| Nzbin | $N_z$ | Number of redshift bins |
| snzmode | nz_read_from_files | File mode |
| nzfile | $f_1$, $f_2$, ..., $f_{\rm Nzbin}$ | File names. See Appendix A.1.5 for the file syntax. |

Table 9: Basic, common part of the configuration file

| | | |
|---|---|---|
| version | double | Config file version. Upwards compatibility (config file version > CosmoPMC version) cannot be guaranteed. Downwards compatibility (config file version < CosmoPMC version) is most likely ensured. |
| | | Parameter section |
| npar | integer | Number of parameters |
| n_ded | integer | Number of deduced parameters. The deduced parameters are not sampled but deduced from the other parameters and written to the output files as well |
| spar | string | Parameterisation type, necessary for the wrapping into the individual posterior parameters and for plotting, see Table 12 for possible parameters |
| min | npar+n_ded doubles | Parameter minima |
| max | npar+n_ded doubles | Parameter maxima |
| | | Data section |
| ndata | integer | Number of data sets |
| sdata | string | Data set 1 |
| | | ⋮ |
| sdata | string | Data set ndata |
| | | Prior section |
| sprior | string | Prior file name ("-" for no prior) |
| [nprior | integer | If sprior ≠ "-": Number of parameters to which prior applies] |
| [indprior | npar × {0, 1} | If sprior ≠ "-": Indicator flags for prior parameters] |

Table 10: PMC part of the configuration file

| | | |
|---|---|---|
| `nsample` | integer | Sample size per iteration |
| `niter` | integer | Number of iterations |
| `fsfinal` | integer | Sample size of final iteration is `fsfinal` $\times$ `nsample` |
| `niter` | integer | Number of iterations (importance runs) |
| `nclipw` | integer | The `nclipw` points with the largest weights are discarded |
| | | **Proposal section** |
| `df` | double | Degrees of freedom (`df`=-1 is Gaussian, `df`=3 is 'typical' Student-t) |
| `ncomp` | integer | Number of components |
| `sdead_comp` | string | One of 'bury', 'revive' |
| `sinitial` | string | Proposal type (one of `fisher_rshift, fisher_eigen, file, random_position`) |
| `fshift`[a] | double | Random shift from ML point $\sim U(-r, r)$; $r$ = `fshift/(max-min)` |
| `fvar`[a] | double | Random multiplier of Fisher matrix |
| `prop_ini_name`[b] | string | File name of initial proposal |
| `fmin`[c] | double | Components have variance $\sim U(\texttt{fmin}, (\texttt{max} - \texttt{min})/2)$ |
| | | **Histogram section** |
| `nbinhist` | integer | Number of density histogram bins |

Table 11: Data-specific entries in the configuration file's data section

| Weak gravitational lensing | Lensing | |
|---|---|---|
| slensdata | string | Data type, one of xipm, xip, xim, map2poly, map2gauss, gsqr, decomp_eb, pkappa, map3gauss, map3gauss_diag, map2gauss_map3gauss, map2gauss_map3gauss_diag, decomp_eb_map3gauss, decomp_eb_map3gauss_diag |
| sdecomp_eb_filter[a] | string | One of FK10_SN, FK10_FoM_eta10, FK10_FoM_eta50, COSEBIs_log |
| th_min[b] | double | Minimum angular scale |
| th_max[b] | double | Maximum angular scale |
| path[b] | double | Path to COSEBIs files |
| sformat | string | Data format of angular scales, one of angle_center, angle_mean, angle_wlinear, angle_wquadr |
| a1[c] | double | Linear weight |
| a2[c] | double | Quadratic weight, $w$ = a1 · $\theta$/arcmin + a2 · ($\theta$/arcmin)$^2$ |
| datname | string | Data file name |
| scov_scaling | string | One of cov_const, cov_ESH09 |
| covname | string | Covariance file name |
| covname_M[d] | string | Covariance mixed term file name |
| covname_D[d] | string | Covariance shot-noise term file name |
| corr_invcov | double | Correction factor for inverse covariance ML estimate, see Hartlap et al. (2007) |
| Nexclude | integer | Number of redshift bin pairs to be excluded from analysis |
| exclude[e] | Nexclude integers | Indices of redshift pairs to be excluded |
| model_file | string | Parameter file name, e.g. cosmo_lens |
| sspecial | string | Additional prior, one of none (recommended), unity, de_conservative |

| Supernovae type Ia | SNIa | |
|---|---|---|
| datname | string | Data file name |
| datformat | string | Data format, SNLS_firstyear |
| schi2mode | string | $\chi^2$ and distance modulus estimator type (one of chi2_simple, chi2_Theta2_denom_fixed, chi2_betaz, chi2_dust, chi2_residual) |
| Theta2_denom[a] | 2 doubles | Fixed $\alpha, \beta$ in $\chi^2$-denominator |
| zAV_name[b] | string | File with $A_V(z)$ table |
| datname_beta_d[b] | string | Prior file (mvdens format) on $\beta_{\rm d}$ ("-" if none) |
| add_logdetCov | integer | 1 if 0.5 log det Cov is to be added to log-likelihood, 0 if not (recommended; see Sect. 4.3) |
| model_file | string | Parameter file name, e.g. cosmo_SN |
| sspecial | string | Additional prior, one of none (recommended), unity, de_conservative |

Table 11: Data-specific entries in the configuration file's data section (continued).

| CMB anisotropies | CMB | |
|---|---|---|
| scamb_path | string | /path/to/scamb |
| data_path | string | /path/to/wmap-data. This path should contain the directory data with subdirectories healpix_data, highl, lowlP, lowlP |
| Cl_SZ_file | string | File with SZ correction angular power spectrum ("-" if none) |
| lmax | integer | Maximum $\ell$ for angular power spectrum |
| accurate | 0\|1 | Accurate reionisation and polarisation calculations in camb |
| model_file | string | Parameter file name, e.g. cosmoDP.par |
| sspecial | string | Additional prior, one of none (recommended), unity, de_conservative |

| WMAP distance priors | CMBDistPrior | |
|---|---|---|
| datname | string | Data (ML point and inverse covariance) file |
| model_file | string | Parameter file name, e.g. cosmo_lens.par |
| sspecial | string | Additional prior, one of none (recommended), unity, de_conservative |

| Galaxy clustering (HOD) | | GalCorr |
|---|---|---|
| shalodata | string | Data type, woftheta |
| shalomode | string | $\chi^2$ type, one of galcorr_var, galcorr_cov, galcorr_log |
| datname | string | Data (+variance) file name |
| covname[a] | string | Covariance file name |
| corr_invcov | double | Correction factor for inverse covariance ML estimate, see Hartlap et al. (2007) |
| delta | double | Power-law slope $\delta$, for integral constraint |
| intconst | double | Integral constant $C$ |
| area | double | Area [deg$^2$] |
| sngal_fit_type | string | Likelihood type, inclusion of galaxy number. One of ngal_lin_fit, ngal_log_fit, ngal_no_fit, ngal_lin_fit_only |
| ngal[b] | double | Number of observed galaxies |
| ngalerr[b] | double | Error on the number of observed galaxies |
| model_file | string | Parameter file name |
| sspecial | string | Not used for HOD, set to none |

Table 11: Data-specific entries in the configuration file's data section (continued).

| Baryonic acoustic oscillations | | BAO |
|---|---|---|
| smethod | string | BAO method, one of distance_A, distance_d_z |
| datname | string | Data + covariance file name (mvdens format) |
| model_file | string | Parameter file name, e.g. cosmoDP.par |
| sspecial | string | Additional prior, one of none (recommended), unity, de_conservative |

Table 12 contains a list of input parameters, which can be given as strings to the spar key in the config file.

Table 12: Input parameters

| Name | Symbol | Description |
|---|---|---|
| Basic cosmology | (some of them given in cosmo.par) | |
| Omega_m | $\Omega_\mathrm{m}$ | Matter density, cold dark matter + baryons |
| omega_m | $\omega_\mathrm{m}$ | |
| Omega_b | $\Omega_\mathrm{b}$ | Baryon density |
| omega_b | $\omega_\mathrm{b}$ | |
| 100_omega_b | $100 \times \omega_\mathrm{b}$ | |
| Omega_de | $\Omega_\mathrm{de}$ | Dark-energy density (if $w = -1$, corresponds to $\Omega_\Lambda$ |
| omega_de | $\omega_\mathrm{de}$ | |
| Omega_nu_mass | $\Omega_{\nu,\mathrm{mass}}$ | Massive-neutrino density (so far only for CMB) |
| omega_nu_mass | $\omega_{\nu,\mathrm{mass}}$ | |
| Omega_c | $\Omega_\mathrm{c}$ | Cold dark matter |
| omega_c | $\omega_\mathrm{c}$ | |
| Omega_K | $\Omega_K$ | Curvature density parameter |
| omega_K | $\omega_K$ | |
| w0_de | $w_0$ | Dark-energy equation-of-state parameter (constant term) |
| w1_de | $w_1$ | Dark-energy equation-of-state parameter (linear term, see sde_param) |
| h_100 | $h$ | Dimensionless Hubble parameter |
| N_eff_nu_mass | $N_{\mathrm{eff},\nu,\mathrm{mass}}$ | Effective number of massive neutrinos (so far only for CMB) |
| sigma_8 | $\sigma_8$ | Power-spectrum normalisation at small scales |
| Delta_2_R | $\Delta^2_\mathcal{R}$ | Power-spectrum normalization at large scales (CMB) |
| n_spec | $n_\mathrm{s}$ | Scalar power-spectrum index |
| alpha_s | $\alpha_\mathrm{s}$ | Running spectral index (so far only for CMB) |
| n_t | $n_\mathrm{t}$ | Tensor power-spectrum index |
| r | $r$ | Tensor to scalar ratio |
| ln_r | $\ln r$ | |
| tau | $\tau$ | Optical depth for reionisation |
| A_SZ | $A_\mathrm{SZ}$ | SZ-power spectrum amplitude |

Table 12: Input parameters (continued)

| SNIa-specific (some of them given in cosmo_SN.par) | | |
| --- | --- | --- |
| M | $M - \log_{10} h_{70}$ | Universal SNIa magnitude |
| alpha | $\alpha$ | Linear response factor to stretch |
| beta | $\beta$ | Linear response factor to color |
| beta_z | $\beta_z$ | Redshift-dependent linear response to color |
| beta_d | $\beta_{\rm d}$ | Linear response to the color component due to intergalactic dust |

| Galaxy-clustering-specific (some of them given in halomodel.par) | | |
| --- | --- | --- |
| M_min | $M_{\rm min}$ | Minimum halo mass for central galaxies $[M_\odot h^{-1}]$ |
| log10_M_min | $\log_{10} M_{\rm min}/(M_\odot h^{-1})$ | |
| M_1 | $M_1$ | Scale mass for satellite galaxies $[M_\odot h^{-1}]$ |
| log10_M_1 | $\log_{10}[M_1/(M_\odot h^{-1})]$ | |
| M_0 | $M_0$ | Minimum halo mass for satellite galaxies $[M_\odot h^{-1}]$ |
| log10_M_0 | $\log_{10} M_0/(M_\odot h^{-1})$ | |
| sigma_log_M | $\sigma_{\log M}$ | Dispersion for central galaxies |
| alpha_halo | $\alpha_{\rm h}$ | Slope of satellite occupation distribution |
| M_halo_av* | $\langle M_{\rm h} \rangle$ | Average halo mass $[M_\odot h^{-1}]$ |
| log10_M_halo_av* | $log_{10}\langle M_{\rm h}/(M_\odot h^{-1})\rangle$ | |
| b_halo_av* | $\langle b_{\rm h} \rangle$ | Average halo bias |
| N_gal_av* | $\langle N_{\rm g} \rangle$ | Average galaxy number per halo |
| fr_sat* | $f_{\rm s}$ | Fraction of satellite galaxies to total |
| ngal_den* | $n_{\rm g}$ | Comoving galaxy number density $[{\rm Mpc}^{-3} h^3]$ |
| log10ngal_den* | $\log_{10} n_{\rm g}$ | |

# 6. Post-processing and auxiliary programs

All scripts described in this section are located in $COSMOPMC/bin.

## 6.1. Plotting and nice printing

### 6.1.1. Posterior marginal plots

Marginals in 1d and 2d can be plotted in two ways, using (1) plot_contour2d.pl or (2) plot_confidence.R. The first is a perl script calling yorick for plotting, the second is an R script. The second option produces nicer plots in general, in particular, smoothing workes better without producing over-smoothed contours. Further, filled contours with more than one data set are only possible with the R option, yorick can only combine several plots with empty contours. The computation time of the R script is however much longer.

1. plot_contour2d.pl creates 1d and 2d marginals of the posterior, from the histogram files chi2_j and chi2_j_k.

   To smooth 1d and 2d posteriors with a Gaussian, use plot_contour2d.pl -n -g FACTOR. The width of the Gaussian is equal to the box size divided by FACTOR. It is recommended to test the smoothing width FACTOR by setting it to a negative number which causes both smoothed and unsmoothed curves being plotted. This can reveal cases of over-smoothing. If contours have very different width in different dimension, the addition option -C uses the PMC sample covariance (from the file covar+ded.fin) as the covariance for the Gaussian. For the final plot, replace -FACTOR with FACTOR to remove the unsmoothed curves. Remove the option -n to add color shades to the 2d contours.

   The file log_plot contains the last plot command with all options. This can be used to reproduce and modify a plot which has been generated automatically by other scripts, e.g. cosmo_pmc.pl.

2. plot_confidence.R creates 1d and 2d marginals of the posterior, from the re-sample file sample.

   Smoothing is done with a kernel density estimation using the R function kde2d. The kernel width can be set with the option -g. The number of grid points, relevant both for smoothing and filled contours, is set with -N. Use both -i and -j options to only plot the 2D marginals of parameters *iandj* to save computation time.

## 6.2. Mean and confidence intervals

From a "mean" output file, containing parameter means and confidence levels, one can create a ps/pdf file using the command mean2eps.pl.

This is equivalent to the following steps (see also essential_cosmo_pmc_run.pl):

- meanvar2tab.pl creates a table with parameter names and values formatted in TeX-format.
- tab2tex.pl wraps a LaTeX table header around the table.

- txt2tex.pl wraps a LaTeX header around the file.

Example:

```
> meanvar2tab.pl -s 1 -p 2 -e iter_9/mean > mean.tab
> tab2tex.pl -s 1.25 mean.tab > mean_in.tex
> txt2tex.pl mean_in.tex > mean.tex
```

#### 6.2.1. PMC proposal

proposal_mean.pl (proposal_var.pl) creates plots of the proposal component's means (variances) as function of the iteration.

### 6.3. Importance sampling

A PMC simulation file (pmcsim) from an earlier PMC run, corresponding to a sample from posterior $p_1$, can be used to do importance sampling with another posterior $p_2$. For that, simply replace the data section of the earlier config file with the corresponding data section of posterior $p_2$. The command importance_sample creates a new PMC simulation which corresponds to a sample under the posterior product $p_1 \cdot p_2$.

### 6.4. Bayesian evidence, Bayes' factor

evidence.pl calculates and prints the evidence from a PMC simulation file. The same information is printed to the file evidence during a PMC run.

bayes_factor.pl prints Bayes' factor between two PMC runs together with the Jeffrey scale.

evidence_list.pl prints a list of evidences for a number of PMC runs.

### 6.5. Reparameterisation

remap.sh swaps and removes parameters from a MCMC or PMC run. The histogram files, mean and covariances are remapped. This is useful if different runs are to be reduced to a common parameter set for comparison or joint plotting. The removal of parameters is equivalent to marginalisation over the corresponding parameter subspace.

For example, suppose there is a SNIa run in directory Sn, and a lensing run in Lensing. SNIa has the following parameters:

```
Omegam Omegade w0de M alpha beta
```

Lensing has the parameters:

```
Omegam sigma8 w0de Omegade h100
```

In Sn, create the file remap.dat with the line

```
0 1 2
```

In Lensing, create the file remap.dat with the line

```
0 3 2
```

In both directories run the command

```
> remap.sh -i iter_<niter-1>
```

which creates sub-directories remap containing symbolic links and/or copies of histogram files to/from iter_{niter-1}, mean, covariance files and updated configuration files.

To create joint marginal plots, simply run

```
> plot_contours2d.pl -c /path/to/Sn/remap/config_pmc -n /path/to/Sn/remap
/path/to/Lensing/remap
```

New parameters, sampled from a flat or Gaussian distribution, can be added using add_par_from_prior.pl.

## 6.6. Analysis

### 6.6.1. mvdens/mix_mvdens format utilities

See Sect. A.3 for a description of the mvdens and mix_mvdens formats.

fisher_to_meanvar.pl reads a mvdens file, inverts the covariance matrix and prints the mean and variance.

corr_coeff.sh reads a mvdens or block matrix file and prints the correlation matrix of the covariance.

diag_mvdens.pl replaces the covariance by its diagonal.

add_par_to_mvdens.pl adds parameters to a mvdens file. Useful, if CosmoPMC is run with additional parameters, and the initial proposal is chosen from a previous run with the reduced parameter set.

### 6.6.2. PMC simulation/MCM chain utilities

sample2fixpar.pl reads a sample file and fixes a parameter by cutting off all points outside a given (narrow) range.

### 6.6.3. PMC proposal diagnostics

neff_proposal.pl calculates the effective number of components (eq. 5). It is the same quantity which is printed to the file enc.

# 7. Using and modifying the code

## 7.1. Modifying the existing code

Note: Code to be used with MPI should not contain global variables and static variables.

## 7.2. Creating a new module

In this section, the steps required to add a new cosmology module to CosmoPMC are described.

1. Create the directory newmodule and create (or copy) files with the necessary code to deal with the data and likelihood. Include files (*.h) should be in newmodule/include, source files (*.c) in newmodule/src. Edit the (or create a new) Makefile (in newmodule) and add the rules libnewmodule.so, libnewmodule.dylib and libnewmodule.a as well as the rule clean.
2. In wrappers/include/types.h:

   Define a new data type by extending the enumeration data_t. Add the corresponding string (for identification of the module in the configuration file) in the macro sdata_t(i), and increase Ndata_t by one.
3. In wrappers/include/all_wrappers.h:

   Add the line

   ```
   #include "newmodule.h"
   ```
4. In tools/include/par.h:

   If necessary, add new parameter types (p_newparameter) to enumeration par_t, add the corresponding identifier strings to the macro spar_t, and increase Npar_t by one.

   Optional: Add the parameter name and syntax for different programs (e.g. gnuplot, yorick, TEX) to bin/spar.txt.

5. In wrappers/src/wrappers.c:

   Add the corresponding case to the 'switch' instruction in the function `init_func_t`. This function sets the data type.

6. Create the files wrappers/include/newmodule.h and wrappers/src/newmodule.c. (Those files need to have different names than the files in newmodule/{src,include}.) Write the following functions:

   a) `init_function_newmodule`
   b) `read_from_config_newmodule`
   c) `init_newmodule`
   d) `likeli_newmodule` (returning log $L$)
   e) `special_newmodule` (optional)
   f) `print_newmodule` (optional)

   To see what these functions are supposed to do, have a look at already existing modules, e.g. in bao.c.

7. In Makefile.main:

   a) In the section "Additional directories", define the path to the new module's directory as

      `NEWMODULE = $(COSMOPMC)/newmodule`

   b) In the section "Libraries", define the library of the new module as

      `LIBNEWMODULE = libnewmodule.$(EXT)`

   c) In the section "Combined cosmo include and linker flags", add the following flags:
      `-I$(NEWMODULE)/include` to the variable `IINCDIRS`
      `-L$(NEWMODULE)` to `LLIBDIRS`
      `-lnewmodule` to `LLIBS`.

8. In exec/Makefile:

   Define the new rule:

```
$(LIBNEWMODULE):
        cd $(NEWMODULE) && $(MAKE) $@
```

   (The second line has to start with a <TAB> and *not* with spaces.)

9. Optional: Extend newdir_pmc.sh.

## 7.3. Error passing system

Most of the situations where an error occurs are intercepted by the program. In such a case, a variable `*err` of type `error*` is set via the macros

```
*err = addError(error_type, "message", *err, __LINE__);
```

or

```
*err = addErrorVA(error_type, "formatted message", *err,
                  __LINE__, VA_LIST);
```

printing the current line and function in the code, a message and the error type (negative integer). With

```
testErrorRet(test, error_type, "message", *err, __LINE__,
             return_value);
```

or

```
testErrorRetVA(test, error_type, "formatted message", *err,
               __LINE__, return_value, VA_LIST);
```

a conditional error is produced if the (Boolean) expression `test` is true. The error is transported up the stack to the calling function with the macro

```
forwardError(*err, __LINE__, return_value);
```

Omit `return_value` in case of a void function. This can be used as diagnostics even for errors deep in the hierarchy of functions.

During the calculation of the importance weights, any error is intercepted and the corresponding point does not contribute to the final sample. See Sect. 2 for more details. Therefore, in the routines which calculate the importance weights, the following is used:

```
forwardErrorNoReturn(*err, __LINE__, return_value);
ParameterErrorVerb(*err, param, quiet, ndim);
```

In case of an error, the first line forwards the error but does not return from the current routine. The second line prints the `ndim`-dimensional parameter `param` to stderr (if `quiet!=1`) and purges the error.

To exit on an error, use

```
quitOnError(*err, __LINE__, FILE)
```

This is usually done only from the main program.

More macros and functions regarding error communication and handling can be found in the files `errorlist.h, errorlist.c` which are part of PMCLIB.

## Acknowledgements

CosmoPMC was developed with support of the CNRS ANR ECOSSTAT, contract number ANR-05-BLAN-0283- 04 ANR ECOSSTAT.

We thank P. Astier, F. Beaujean, J. Guy, L. Fu, A. Halkola, J. Hartlap, B. Joachimi, J. Lui, K. Markovič, P. Schneider, F. Simpson, R. E. Smith, M. Takada and I. Tereno for discussions and insights which helped to develop the cosmology code.

We thank L. Fu for helping with and testing the lensing E-/B-mode decomposition and the third-order lensing code.

We acknowledge R. .E. Smith and J. A. Peacock for making public their code `halofit`[7], which we implemented into CosmoPMC.

The people from the Coyote project[8] are thanked for making their code public. An adapted version of their emulator is part of this code.

The following people are thanked for providing data or simulations:

M. Kowalski for the Union data (Kowalski et al. 2008), J. Hartlap and S. Hilbert for the Lensing covariance, computed using ray-tracing through the Millennium Simulation (Schrabback et al. 2010).

[7] `http://www.roe.ac.uk/~jap/haloes`

[8] `http://www.lanl.gov/projects/cosmology/CosmicEmu`

## PMC references

| | |
|---|---|
| Introductory papers on PMC | |
| Cappé et al. (2004) | Population Monte Carlo |
| Cappé et al. (2008) | Adaptive importance sampling in general mixture classes |
| Comparison of sampling methods including PMC | |
| Robert & Wraith (2009) | Computational methods for Bayesian model choice |
| Main papers on CosmoPMC | |
| Wraith et al. (2009) | Estimation of cosmological parameters using adaptive importance sampling |
| Kilbinger et al. (2010) | Bayesian model comparison in cosmology with Population Monte Carlo |
| PMC applied to cosmological data | |
| Schrabback et al. (2010) | Evidence of the accelerated expansion of the Universe from weak lensing tomography with COSMOS |
| Ménard et al. (2010) | On the impact of intergalactic dust on cosmology with Type Ia supernovae |
| Benabed et al. (2009) | TEASING: a fast and accurate approximation for the low multipole likelihood of the cosmic microwave background temperature |
| Coupon et al. (2012) | Galaxy clustering in the CFHTLS-Wide: the changing relationship between galaxies and haloes since $z \sim 1.2\star$ |
| Kilbinger et al. (2012) | CFHTLenS: Combined probe cosmological model comparison using 2D weak gravitational lensing |
| Benjamin et al. (2012) | CFHTLenS tomographic weak lensing: Quantifying accurate redshift distributions |
| Simpson et al. (2012) | CFHTLenS: Testing the Laws of Gravity with Tomographic Weak Lensing and Redshift Space Distortions |
| Other publications which use PMC | |
| Joachimi & Taylor (2011) | Forecasts of non-Gaussian parameter spaces using Box-Cox transformations |
| Beaujean et al. (2012) | Bayesian fit of exclusive $b \to s\bar{\ell}\ell$ decays: the standard model operator basis |

# A. File formats

## A.1. Data files

### A.1.1. Lensing

For all `lensdata_t` types, the data format is the same. Each line contains the data for a given angular scale and (arbitrary many) redshift bin pair combinations.

The angular scales are defined as follows. For `lensformat = angle_center`, the fist column contains the angular bin center in arc minutes. For the cases `lensformat = angle_mean`,

`angle_wlinear` and `angle_wquadr`, first two columns specify the lower and upper end of the angular bin.

Following the angular information are the data. For $N_z$ redshift bins, $N_z(N_z+1)/2$ columns specify all pair combinations $(ij)_{i\leq j}$ in lexical order, that is $(11)(12)(13)\ldots(1N_z)(22)(23)\ldots(N_zN_z)$.

Note that for `lensdata = xipm` the first $N_\theta$ lines of the data file contain $\xi_+$ for $N_\theta$ angular scales, the last $N_\theta$ lines contain $\xi_-$, where the angular scales (first or first two columns) are identical in both halfs.

The covariance matrix is in block format: It consists of $N$ lines and $N$ columns, where $N = N_\mathrm{s}N_z(N_z+1)/2$ is the length of the data. Usually, $N_\mathrm{s}$ is the number of measured angular scales, $N_\theta$, unless there is more than one data point per scale (e.g. for `lensdata = xipm`, $N_\mathrm{s} = 2N_\theta$).

A matrix element $C_{ij}$ equals $\langle d_i d_j\rangle - \langle d_i\rangle\langle d_j\rangle$, where $d_i$ is the $i^\mathrm{th}$ data point. In the counting over angular scale and redshift, the former varies faster than the latter[9]. For example, with two redshift bins and three angular scales, the element $C_{77}$ is the data variance for the redshift pair (11) and angular scale $\theta_1$ (starting counting at zero). Or, in other words, the covariance matrix consists of $N_z(N_z+1)/2$ block sub-matrices, each of size $N_\mathrm{s}\times N_\mathrm{s}$. Each sub-matrix corresponds to one redshift bin combination. It is therefore easy to exclude some redshift bins, by (1) setting the diagonal of a sub-matrix to a very high value, and (2) setting the off-diagonal to zero (see the `Nexclude` parameter in the config file, Table 11).

### A.1.2. SNIa

The SNIa data file in `SN_SALT` format starts with the following two lines:

```
@INTRINSIC_DISPERSION double
@PECULIAR_VELOCITY double
```

The peculiar velocity value is in units of km/s. This is followed by a list of supernovae, one object on each line as follows:

$$\text{name} \;\; z \;\; m \;\; s \;\; c \;\; <m^2> \;\; <s^2> \;\; <c^2> \;\; <ms> \;\; <mc> \;\; <sc>$$

### A.1.3. BAO

The BAO distance measures are modeled as Gaussian variables, the data files are in `mvdens` format (see Sect. A.3). In the same file, following the `mvdens` data, there is a list of redshifts, corresponding to where the distances are measured.

[9] This was wrongly stated here until version 1.01.

### A.1.4. CMB

The CMB data for WMAP are the ones released by the WMAP team. They are not included in CosmoPMC and can be obtained e.g. from the LAMBDA site[10].

The SZ correction power spectrum file has two columns in each row containing $\ell$ and $C_\ell$, respectively. The first line has to start with $\ell = 2$.

The CMB distance priors (Komatsu et al. 2009) are given in `mvdens` format.

### A.1.5. Redshift distribution

The first line of a file describing a the redshift distribution for a redshift bin contains the type, see Sect. 4.7,

```
# nofz
```

This is followed by the list of parameter values, in the order given in Table 3. Each parameter value has to be in a new line, with the exeption of the histogram, `nofz = single`. In that case, the parameter lines are as follows:

$$\begin{array}{cc} z_0 & N_0 \\ z_1 & N_1 \\ \ldots & \\ z_{n-1} & N_{n-1} \\ z_n & 0 \end{array}$$

$N_i$ is the number of galaxies in the bin $[z_i; z_{i+1}]$. The last line denotes the upper limit of the last histogram bin $z_n = z_{\rm max}$, followed by a zero. For `nofz = single`, the file has to contain two identical lines with the value of $z_0$ in each line.

## A.2. Output file names

The default names of all output files are defined in `stdnames.h`. Edit this file and to `make clean; make` to set user-defined file names. Note however that some of the pre-processing scripts expect the default names.

## A.3. Multi-variate Gaussian/Student-t (`mvdens`), mixture models (`mix_mvdens`)

The `mvdens` file format is as follows. The first (header) line contains four integers:

[10] `http://lambda.gsfc.nasa.gov`

$$p \quad \nu \quad B \quad c.$$

Here, $p$ is the number of dimensions, $\nu$ the degrees of freedom. For a multi-variate Gaussian, choose $\nu = -1$, and $\nu > 0$ for Student-t. $B$ indicates the number of secondary diagonal of the covariance matrix which are updated during the PMC iterations. For most purposes, $B$ can be set equal to $p$, which corresponds to the whole matrix being updated. Finally, $c$ is 1 if the matrix is Cholesky-decomposed and 0 otherwise.

This is followed by $p$ doubles indicating the mean, followed by $p$ lines with $p$ doubles each, giving the (symmetric) covariance matrix.

Here is an example of a 5-dimensional multi-variate Gaussian (not Cholesky-decomposed):

```
5 -1 5 0
0.38559 -1.5238 19.338 1.3692 -2.4358
0.0053677 -0.025608 0.00066748 -0.0011893 0.00087517
-0.025608 0.16837 -0.0079163 0.0027364 -0.0035709
0.00066748 -0.0079163 0.0011077 0.0010986 -0.00067815
-0.0011893 0.0027364 0.0010986 0.016716 0.0026266
0.00087517 -0.0035709 -0.00067815 0.0026266 0.014881
```

The `mix_mvdens` format has two doubles as the header:

$$D \quad p$$

where $D$ is the number of components of the mixture and `ndim` the dimension. This is followed by $D$ blocks specifying the weights $w_d$ (doubles) and data $m_d$ (in `mvdens` format) of the $D$ multi-variate densities of the mixtures.

$$\begin{array}{c} w_1 \\ m_1 \\ w_2 \\ m_2 \\ \ldots \\ w_D \\ m_D. \end{array}$$

The weights should be normalised, $\sum_{d=1}^{D} w_d = 1$.

In many cases, an `mvdens` file indicates a parameter covariance matrix, for example to be used as Gaussian prior using the config file flag `sprior`. In some cases, the inverse covariance matrix is expected, as in the case of the Fisher matrix.

# B. Syntax of all commands

All following scripts are located in `$COSMOPMC/bin`. All programs (executables) are located in `$COSMOPMC/exec` and linked from `$COSMOPMC/bin` after running `make` in `$COSMOPMC`.

- `add_deduced_halomodel` 11, 22

```
Usage: add_deduced_halomodel [OPTIONS] PSIM [PAR_1 [PAR_2 [...]]]
OPTIONS:
  -c CONFIG        Configuration file (default: config_pmc)
  -o OUTNAME       Ouput pmcsim name (default: psim+ded)
  PSIM             pmc simulation file (pmcsim_iter)
  PAR_i            String for deduced parameter #i. If not given, deduced
                    parameters are read from the config file (default)
```

- `add_par_from_prior.pl` 37

```
Usage: add_par_from_prior.pl [OPTIONS] sample
Adds a new random parameter to a PMC sample file, drawn under a distribution
OPTIONS:
  -o OUT      Output sample file OUT (default: '<sample>.out'
  -p DIST     Prior distribution, DIST one of 'Flat' (default), 'Gauss'
  -P ARG      Prior arguments (white-spaced list if more than one). For DIST =
               Flat:  ARG = 'min max' (defaut '-1 1')
               Gauss: ARG = 'mean sigma'
  -C COL      Column COL of new parameter (default: last)
  -s STR      Name string STR of new parameter
  -h          This message
```

- `add_par_to_mvdens.pl` 37

```
add_par_to_mvdens.pl (MIX)MVDENS [OPTIONS]
Adds a parameter to a (mix)mvdens file (e.g. Fisher matrix, PMC proposal)
OPTIONS:
  -c COL     Adds parameter in column and row COL (default: last column)
  -m VAL     Parameter mean VAL (default 0)
  -v VAL     Parameter variance VAL (default 1)
  -x         File is in 'mixmvdens' format
   FILE      File name
  -h         This message
```

- `bayes_factor.pl` 36

```
Usage: bayes_factor.pl [OPTIONS] DIR1 DIR2
Calculates the Bayes factor between models. The corresponding
 evidence files (from PMC) have to be in the directories DIR1 and DIR2
OPTIONS:
  -i 'ITER1 [ITER2]'  Use iteration ITER1 for DIR1 and ITER2 for DIR2
                       (default: all iterations)
  -f 'EVI1 [EVI2]'    Use files DIR1/EVI1 and DIR2/EVI2 (default: 'evidence')
  -s                  Short output, last iteration only
  -l                  Laplace approx. from Fisher matrix (denoted with iter=-1)
  -h                  This message
```

- `cl_one_sided` 10

```
Usage: cl_one_sided [OPTIONS] sample
OPTIONS:
  -c CONFIG        Configuration file (default: config_pmc)
  -i INDEX         Parameter index
  -d DIR           Direction (DIR=+1,-1)
  -v VALUE         Starting value
  -w WHICH         WHICH=0: 68%,95%,99.7% c.l. (default)
                   WHICH=1: 68%,90%,95% c.l.
  sample           PMC sample file
  The options -i INDEX, -d DIR and -v VALUE are required
```

- config_pmc_to_max_and_fish.pl 7, 26

```
Usage: config_pmc_to_max_and_fish.pl [OPTIONS]
OPTIONS:
  -M            Create config file for maximum search (max_post)
  -F            Create config file for Fisher matrix (go_fishing)
  -c CONFIG     Input PMC config file CONFIG (default: 'config_pmc')
  -r            Random starting point (for maximum search)
  -f FID        Fiducial starting point FID. FID is a white-space
                 separated list in quotes, e.g. '0.25 0.75'
  -p FILE       Fidcucial parameter from FILE (e.g. 'maxlogP')
  -t TOLERANCE  Tolerance for maximum-search (default: 0.01)
  -d            Calculate only diagonal of Fisher matrix (go_fishing)
  -h            This message
One of '-M' or '-F' is obligatory
The default starting point for maximum search is (max-min)/2
For Fisher matrix ('-F'), a fiducial parameter has to be indicated with '-f FID'
 or '-p FILE'
```

- corr_coeff.sh 37

```
Usage: corr_coeff filename [mvdens|block]
```

- cosmo_mcmc

```
Usage: cosmo_mcmc [OPTIONS]
OPTIONS:
  -c CONFIG        Configuration file (default: config_mcmc)
  -s SEED          Use SEED for random number generator. If SEED=-1 (default)
                    the current time is used as seed.
  -h               This message
```

- cosmo_pmc 12

```
Usage: cosmo_pmc [OPTIONS]
OPTIONS:
  -c CONFIG        Configuration file (default: 'config_pmc')
  -s SEED          Use SEED for random number generator. If SEED=-1 (default)
                    the current time is used as seed.
  -q               Quiet mode
  -h               This message
```

- cosmo_pmc.pl 5, 7, 35

```
Usage: cosmo_pmc.pl [OPTIONS]
```

```
OPTIONS:
   -n NCPU               Run PMC in parallel on NPCU cpus using 'mpirun' (default: 1)
   -c CONFIG             Configuration file for PMC (default: config_pmc)
   -f FID                Fiducial starting point FID. FID is a white-space
                          separated list in quotes, e.g. '0.25 0.75'
   -r                    Random starting point for maximum search
                          (default: (max-min)/2)
   -m [c|a]              Maximum-search method: 'c' (cg), 'a' (amoeba)
   -d                    Calculate only diagonal of Fisher matrix
   -D                    Do not force Fisher matrix F to be positiv. If F is negative,
                          script exits with an error
   -a                    Adaptive numerical differentiation for Fisher matrix
   -s SEED               Use SEED for random number generator. If SEED=-1 (default)
                          the current time is used as seed.
   -S [M|F]              Stops after maximum search ('M') or Fisher matrix ('F')
   -A [y|n]              Default answer to all questions on stdin
   -P PATH               Use PATH as CosmoPMC directory (default: environment
                          variable $COSMOPMC)
   -e                    Create 'essential' plots
   -p PRO                Plotting scripts: 'y' (yorick; default), 'R' (R) or 'n' (none)
                          Combinations of letters are possible, e.g. 'yR'
   -M MULT               Output sample MULT times input (default 1).
                          Valid if plotting script is 'R'
   -O OPT                Pass options OPT to 'plot_contour2d.pl'
   -q                    Quiet mode
   -h                    This message
```

- diag_mvdens.pl 37

```
Usage: diag_mvdens.pl IN
    Prints the mvdens file 'IN' with the covariance replaced by its diagonal.
```

- essential_cosmo_pmc_run.pl 35

```
Usage: essential_cosmo_pmc.pl [OPTIONS]
OPTIONS:
    -c CONFIG      Uses config file CONFIG (default: 'config_pmc')
    -P PATH        Use PATH as CosmoPMC directory (default: environment
                    variable $COSMOPMC)
    -k             Keep temporary files
    -v             Verbose
    -h             This message
```

- evidence.pl 36

```
Usage: evidence.pl [OPTIONS] SAMPLE
OPTIONS:
   -h          This message
SAMPLE         PMC sample file
```

- evidence_list.pl 36

```
Usage: evidence_list.pl [OPTIONS] DIR1 [DIR2 [...]]
OPTIONS:
  -r N             Subtract log(E) from DIRN (default: no subtraction)
```

```
                  For N=-1 subtract log(E_min)
   -k KEY         Use KEY (string list) instead of
                  directory names (default)
   -s SEP         Use SEP as input separator for KEY list
   -S SEP         Use SEP as output separator
                  (default for both: white-space)
   -n             Write number of model parameters
   -L             Use Laplace approximation (reading file 'evidence_fisher')
   -h             This message
```

- fisher_to_meanvar.pl 37

```
fisher_to_meanvar.pl [OPTIONS] file
OPTIONS:
    -n              No inverse
    -m              Marginal errors (don't invert matrix)
    -x              mixmvdens format (default: mvdens format)
    -k              Keep temporary file 'fishtmp.i'
    -h              This message
Options '-m' and '-n' exclude each other
```

- get_spar.pl

```
Usage: get_spar.pl [OPTIONS] LANG [PAR1 [PAR2 [...]]]
OPTIONS:
   -c CONFIG          Configuration file ONFIG (default 'config_pmc')
   -i INDEX           Returns only par[INDEX]
   -P PATH            Use PATH as CosmoPMC directory (default: environment
                       variable $COSMOPMC)
   -p                 Print 'p<i> for unknown parameters instead of input string
   LANG               One of 'yorick', 'gnuplot', 'TeX', 'R'.
                       More languages can be defined in spar.txt
   PAR1 ...           Prameter strings
```

- go_fishing 7, 12, 12

```
Usage: go_fishing [OPTIONS]
OPTIONS:
  -c CONFIG        Configuration file (default: config_fish)
  -a               Adaptive numerical differentiation (default: fixed difference)
  -f               Force positive Fisher matrix
  -q               Quiet mode
  -h               This message
Run in parallel on NP cpu's: 'mpirun -np NP go_fishing [OPTIONS]
```

- haloplot 22

```
Usage: haloplot log10(M_min) log10(M1) log10(M0) sigma_log_M alpha_halo
         halomodel.par [OPTIONS]
Outputs HOD-derived quantities
OPTIONS:
  -o OUT          Output file name
  -t TYPE         Output type, TYPE in [wtheta, wp, xi, xihalo, deltaSig,
                   nofm, halo, pk], default: wtheta
  -nbins          Number of bins
```

```
  -range          Range (linear scale): min,max
  -z Z            Used fixed redshift Z (no w(theta) output)
  -Mhalo log10M   log10(Halo mass) for deltaSig and xihalo (in M_sol/h)
  -c CONFIG       PMC config file, to calculate chi^2
  -h              This message

M_min, M1 and M0 are in units of M_{sol}/h.
```

- `histograms_sample` 11, 11

```
Usage: histograms_sample [OPTIONS] sample
OPTIONS:
  -c CONFIG        Configuration file (default: config_pmc)
  -1               Only 1d histograms
  -2               Only 2d histograms
  sample           PMC sample file
  -h               This message
```

- `importance_sample` 36

```
Usage: importance_sample [OPTIONS] INSAMPLE
Performs an importance run on a PMC sample. Run in
 parallel with MPI (use mpirun)
OPTIONS:
  -c CONFIG        Configuration file (default: config_pmc)
  -o OUTSAMPLE     Output sample name (default: 'insample.out')
  -q               Quiet mode
  -h               This message
  INSAMPLE         Input sample name
```

- `max_post` 7, 12, 12

```
Usage: max_post [OPTIONS]
OPTIONS:
  -c CONFIG        Configuration file (default: config_max)
  -m [c|a|n]       Maximum-search method: 'a' (amoeba, default), 'c' (cg),
                    'n' (none; print posterior for fiducial parameter and exit)
  -t               Test maximum at the end
  -s SEED          Use SEED for random number generator. If SEED=-1 (default)
                    the current time is used as seed.
  -p               Prints the maximum-posterior model to the file 'model_maxlog'
  -q               Quiet mode
  -h               This message
```

- `mean2eps.pl` 35

```
Usage: mean2eps.pl [OPTIONS] MEAN
OPTIONS:
    MEAN           File containing mean and confidence levels (output of
                    'cosmo_pmc' or 'histograms_sample'
    -c CONFIG      Uses config file CONFIG (default: 'config_pmc')
    -P PATH        Use PATH as CosmoPMC directory (default: environment
                    variable $COSMOPMC)
    -o BASE        Outname BASE (default: <MEAN>)
    -v             Verbose
```

```
   -h              This message
```

- `meanvar2tab.pl` 35

```
Usage: meanvar2tab.pl [OPTIONS] file [file2 [...]]

Options:
  -s {123}     68% (1), 95% (2) or 99.7% (3) errors (default = 1)
  -p PREC      Output with PREC digits ('%PREC' format string)
  -e           Error(s) written to PREC significant digits (use -p PREC)
  -c CONFIG    Uses config file CONFIG (default: 'config_pmc')
  -t TITLE     Title (table heading) TITLE is string list with entries according
               to the number of input files
  -S SEP       Use SEP as input separator for TITLE list (default: white space)
  -P PATH      Use PATH as CosmoPMC directory (default: environment
                variable $COSMOPMC)
  -h           This message
```

- `meanvar_sample` 10

```
Usage: meanvar_sample [OPTIONS] sample
OPTIONS:
  -c CONFIG        Configuration file (default: config_pmc)
  -w               Ignore weights (default: weights=first column of sample file)
  -C               Write covariance and inverse covariance to files
  -E               Output evidence
  -h               This message
  sample           PMC sample file
```

- `neff_proposal.pl` 10, 38

```
Usage: neff_proposal.pl PROP
   Calculates the effective number of components for the mix_mvdens file 'PROP'
```

- `newdir_pmc.sh` 5, 39

```
Usage: newdir_pmc.sh [DIR]
Directory DIR (default: read on input) is created.
Links are set to data files in \$COSMOPMC/data.
Parameter files are copied on request from \$COSMOPMC/par_files.
```

- `plot_confidence.R` 6, 11, 35

```
Usage: plot_confidence.R [options]


Options:
-h, --help
Show this help message and exit

-N NGRID, --Ngrid=NGRID
Number of grid points for smoothing (kde2d) (default 100). Use <=30 for
fast-but-dirty plots

-g GSMOOTH, --gsmooth=GSMOOTH
Smoothing kernel width, with respect to box size (default 30). In case of more
```

```
than one sample, use list separated with '_' for more than value

-S, --solid
All contours with solid lines

-w WIDTH, --width=WIDTH
Line width (default 1)

-k, --with_keys
Add key to plots

-K KEYSTRING, --keystring=KEYSTRING
Key strings (separate items with '_')

-L, --no_key_line
Do not add a line to the keys in the legend

-c CONFIG, --config=CONFIG
Config file (default 'config_pmc')

-t TITLE, --title=TITLE
Title string for each panel (default empty)

-i INDEX_I, --index_i=INDEX_I
Only create plots with i-th parameter on x-axis

-j INDEX_J, --index_j=INDEX_J
Only create plots with j-th parameter on y-axis

-s SIGMA, --sigma=SIGMA
Plot SIGMA confidence levels (default 3)

-F COLOR_SCHEME, --color_scheme=COLOR_SCHEME
Color scheme (0, 1; default 0)
```

- plot_contour2d.pl 6, 11, 11, 35

```
Usage: plot_contour2d.pl [OPTIONS] [DIR1 [DIR2 [...]]]

OPTIONS:
  -i NITER       Number of iterations (needed if do_proposal=2)
  -c CONFIG_FILE Configuration file (default: in order config_mcmc, config_pmc)
  -t TITLE       Title string for each panel (default empty)
  -T TITLE       Title string for all_contour2d.{eps|pdf} (default empty)
  -n             No shade
  -w WIDTH       Line width WIDTH (default 4)
  -1 OPT         Add 1d posterior plots. OPT can contain the following letters:
                  m    Plot line at mean position
                  123  Plot line at 68%,95%,99.7 density
```

```
                  t    Write mean and 68% confidence intervals as text
                        (use with 'm' and '1'
                  n    None of the above
  -S              All contours with solid lines
  -s N            Outermost level is N sigma
  -r              Aspect ratio=1, changes plot limits such that dx=dy
  -g FACTOR       Gaussian smoothing of 2d-histograms with variance
                   box-width/|FACTOR|. If FACTOR is negativ, plots
                   unsmoothed histogram in addition (use with '-n').
                   Note: For multiple contours, use a list of values "g1 g2 ..."
  -G FACTOR       Gaussian smoothing of 1d-histograms (default: 2d factor)
  -C              Use covariance (file covar.fin) for Gaussian smoothing
  -N NORM         Normalisation of 1d posterior
                   'm'  Maximum = 1 (default)
                   'i'  Integral over posterior = 1
  -F NUM          Color scheme, NUM=0,1,2
  -k              Add key to plots
  -K "KEY1 [KEY2 [...]]" Key strings (default: directory names)
  -y FS           Font size FS (default 24)
  -o FORMAT       Output file format, FORMAT=eps|pdf (default: eps)
  -b              Writes the chi2 files in block format
  -m PAR          Plots a mark at position PAR (e.g. best-fit). PAR is white-space
                   separated list (use quotes or '\ ', e.g. '0.3 0.8')
  -P PATH         Use PATH as CosmoPMC root directory (default: environment
                   variable $COSMOPMC)
  -q              Run quietly, no verbose
  -h              This message
  DIR1 ...        List of directories containing histogram files (chi2_*_*)
                   Default: DIR1 = '.'
```

- proposal_mean.pl 10, 36

```
Usage: proposal_mean.pl [OPTIONS]
OPTIONS:
  -d DIR          Directory DIR containing the sub-directories 'iter_*'
                   with the proposal files (default '.')
  -c CONFIG       Configuration file CONFIG (default 'DIR/config_pmc')
  -n              No plotting, only creates '.gnu' file
  -i              x- and y-axes inverted
  -I              x- and y-labels on top/right
  -P PATH         Use PATH as CosmoPMC root directory (default: environment
                   variable $COSMOPMC)
  -h              This message
```

- proposal_var.pl 10, 36

```
Usage: proposal_var.pl [OPTIONS]
OPTIONS:
  -d DIR          Directory DIR containing the sub-directories 'iter_*'
                   with the proposal files (default '.')
  -c CONFIG       Configuration file CONFIG (default 'DIR/config_pmc')
  -P PATH         Use PATH as CosmoPMC root directory (default: environment
```

```
                    variable $COSMOPMC)
  -h                This message
```

- remap.sh 36

```
Usage: remap.sh [OPTIONS]
OPTIONS:
   -c CONFIG           Input PMC configuration file (default './config_pmc')
   -i INPUT            Input directory INPUT (default '.')
   -s PMCSIM           Sample/PMC simulation file PMCSIM
   -o OUTPUT           Output directory OUTPUT (default './remap')
   -r REMAP            Remap file REMAP (default './remap.dat')
   -n NPAR             Number of parameters NPAR (default: read from remap file)
   -d N_DED            Number of deduced parameters N_DED (default: 0)
   -h                  This message
```

- sample2fixpar.pl 38

```
Usage: sample2fixpar.pl SAMPLE_IN COL MIN MAX
    SAMPLE_IN          Input sample (PMC simulation or MCM chain)
    COL                Column number of fixed parameter
                        (Note that par #i is in column i+2)
    MIN, MAX           Minimum and maximum values for fixed parameter
```

- tab2tex.pl 35

```
Usage: tab2tex.pl [OPTIONS] file

OPTIONS:
   -a          Produce tex array, not tex table
   -b          Bare output, no table/array header
   -s STRETCH  Set arraystretch to STRETCH
   -m          Add '$' around entries (tex inline math mode)
   -l MODE     Print vertical lines between rows according to MODE;
                a     all lines (default)
                n     no lines
                h     header lines
   -L MODE     Print horizontal lines between columns according to Mode:
                a     all lines (default)
                n     no lines
   -h          This message
```

- test_suite_cosmo_pmc.pl

```
Usage: test_suite_cosmo_pmc.pl [OPTIONS]
OPTIONS:
  -r              Do PMC test runs
  -R              Only do PMC test runs
  -n NCPU         Run PMC in parallel on NCPU cpus (default: 1)
  -c              Include CMB tests
  -P PATH         Use PATH as CosmoPMC root directory (default: environment
                   variable $COSMOPMC)
  -s              Short, without time-taking PMC runs (e.g. Lensing/COSMOS-S10)
  -k              Keep temporary files
  -x              Clean previous run and exit
```

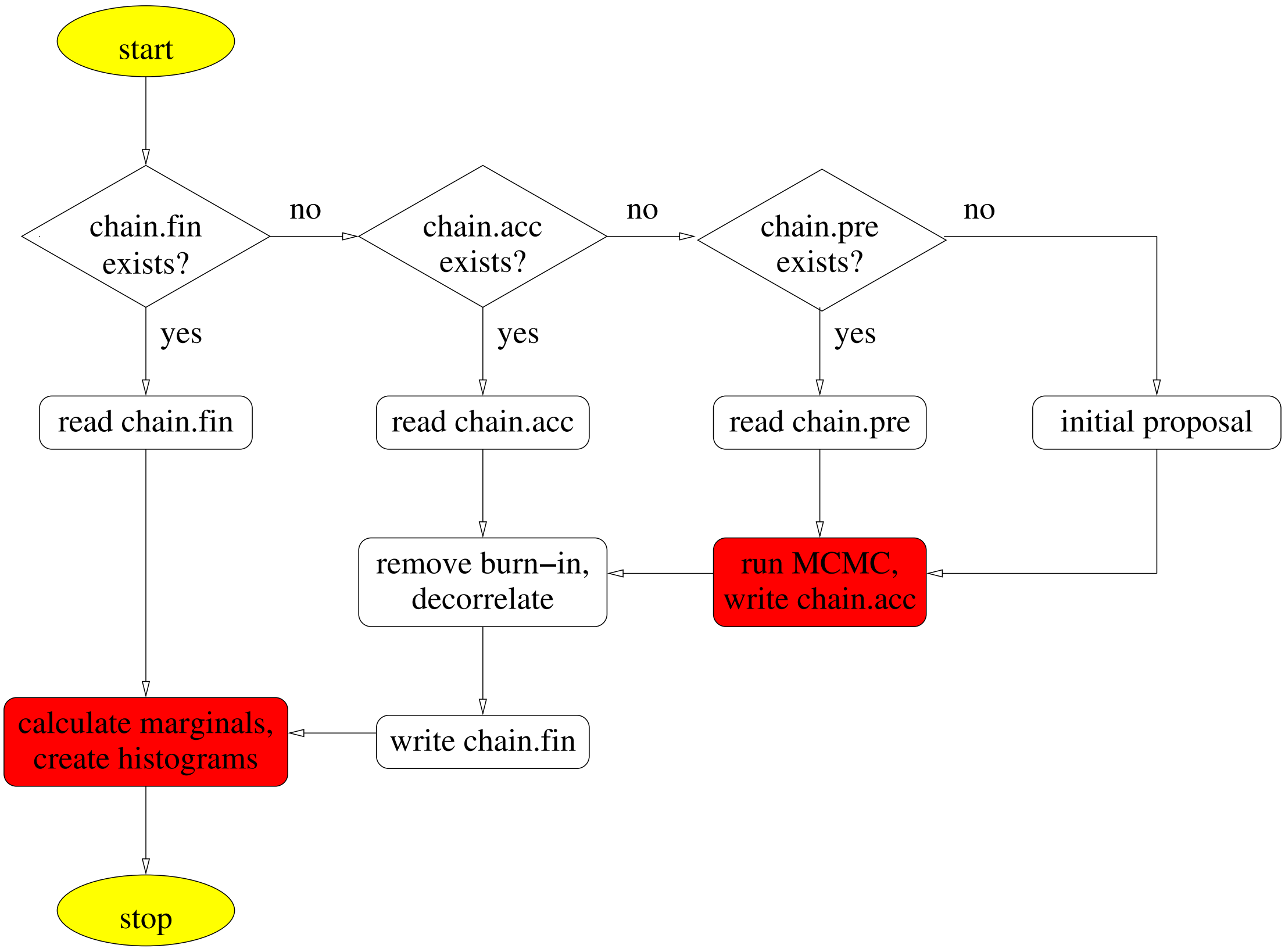

Figure 2: Flow chart of the MCMC implementation.

```
-v            Verbose
-h            This message
```

## C. MCMC

We provide a Metropolis-Hastings Monte-Carlo Markov Chain sampler, which is included in the CosmoPMC package. This MCMC implementation has been used in Wraith et al. (2009) in the comparison with PMC. In the following, we briefly describe our MCMC program.

Table 13: MCMC section of the configuration file

| | | |
|---|---|---|
| nchain | integer | Chain Length |
| ncov | integer | Interval between updates of the proposal covariance |
| fburnin | double | Burn-in phase are the first ncov×ncor points |
| ndecorr | double | De-correlation (thinning-out): one in ndec points is kept in the final chain |
| fudge | double | Proposal covariance is multiplied by fudge$^2$/n_par |
| sinitial | string | Initial proposal type, one of Fisher_inv, Fisher, Fisher, previous, Hessian, Hessian_diag, diag. |
| boxdiv[a] | double | Diagonal of proposal covariance is (max-min)/boxdiv |
| sstart | string | Starting point type, one of ran, fid, min, max, nul |
| fid[b] | npar doubles | Starting parameter |
| | Histogram section | |
| nbinhist | integer | Number of density histogram bins |

[a]only if sinitial = diag
[b]only if sstart = fid

## C.1. MCMC configuration file

## C.2. Proposal and starting point

The proposal for the Metropolis-Hastings algorithm is a multi-variate Gaussian distribution. After choosing an initial proposal, a new proposal can optionally be re-calculated after a number of ncov (accepted) steps. The covariance of this new proposal is the chain covariance from steps up to this point. This proposal is then updated after each ncov accepted steps using all previous accepted points.

There are several options for the initial proposal:

1. **sinitial = diag** A diagonal covariance with width being a fraction of the box size.
2. **sinitial = Fisher** The Hessian at a given point in parameter space. If this point is the maximum-likelihood point, the Hessian corresponds to the Fisher matrix.
3. **sinitial = Fisher_inv** The inverse Hessian/Fisher matrix, e.g. the covariance from a previous chain. This can be useful for ill-conditioned matrices which are difficult to invert numerically.
4. **sinitial = previous** A proposal read from a file, e.g. from a previous MCMC run.

The starting point is either chosen randomly or specified in the config file. The second case might be convenient if the prior volume is very large and a very long burn-in phase is to be avoided. For example, the ML-point or best-fit value from a previous experiment can be chosen Dunkley et al. (2009).

## C.3. Output files

The MCMC output files have the same format as their PMC counterparts (see Sect. 3.3.2).

A complete run of `cosmo_mcmc` produces three files containing the points of the Markov chain:

1. `chain.all` containing all, accepted and rejected, sample points. This is the only chain file will not be read or used in subsequent calls of `cosmo_mcmc`.
2. `chain.acc` containing the accepted points.
3. `chain.fin` containing the accepted points after removal of the burn-in phase and after de-correlating (thinning-out) the chain. The results produced by `cosmo_mcmc` (mean, errors, histograms, covariance) are based on this file.

The chains are `ASCII`-files, in the same format as the PMC sample files. All weights are 1, and the second column contains the log-likelihood (only in `chain.all`.

The parameter mean and confidence intervals are printed to the file `mean`. The names of files containing the histograms and parameter covariances are the same as for PMC.

## C.4. Diagnostics

In general it is not straight-forward to diagnose an MCM chain. There exists tests but no formal proofs for convergence (e.g. Gellman-Rubin), which in addition require very long or multiple chains. We have not implemented such tests in the code. However, there are a few (rather hand-waving) diagnostic tools to check the reliability of an MCMC run.

Firstly, the acceptance rate $\eta$ should be in the range between 15% and 25%. A larger $\eta$ most probably corresponds to a chain which stayed mainly in the high-density region and strongly under-sampled the lower-density posterior regions. In that case the error bars will be underestimated. A very small $\eta$ means probably an under-sampling of the posterior since only few points are accepted. However, this need not cause a bias for the parameters and errors if the chain has been run long enough.

## C.5. Resuming an interrupted run

Sometimes a MCMC run is interrupted before finishing, or one wishes a previous run to be extended, for example because its convergence is doubted. The MCMC program allows to read in and extend a previous chain. To that end, rename the file `chain.acc` into `chain.pre`. The proposal for the resumed run can but need not be calculated from the previous chain (to be controlled in the config file, see Sect.C.2). In the config file, the number of desired sample points has to be larger than the previous chain.